\documentclass[useAMS,usenatbib]{mn2e}
\usepackage{graphicx}
\usepackage[figuresright]{rotating}

\def\ngc{NGC\,6231}
\def\hd{HD\,152}
\def\hda{HD\,152248}
\def\cpd{CPD\,$-$41$^{\circ}$}
\def\sco{Sco\,OB\,1}

\def\bcep{$\beta$\,Cep}

\def\l{$\lambda$\,}
\def\halph{H$\alpha$}

\def\hea{He\,{\sc i}}

\def\kms{km\,s$^{-1}$}
\def\cnts{cnt\,s$^{-1}$}
\def\ergs{erg\,s$^{-1}$}
\def\ergscm{erg\,cm$^{-2}$\,s$^{-1}$}

\def\lg{$\log - \log$}

\def\chandra{\emph{Chandra}}
\def\rosat{\emph{{\sc ROSAT}}}

\def\einst{\emph{{\sc EINSTEIN}}}

\def\xmm{{\sc XMM}\emph{-Newton}}

\def\epic{{\sc EPIC}}
\def\mos{{\sc MOS}}
\def\pn{pn}

\def\epicmos{{\sc EPIC MOS}}
\def\epicpn{{\sc EPIC} pn}

\def\xspec{{\sc xspec}}
\def\mek{{\sc mekal}}
\def\chin{$\chi^2_{\nu}$}
\def\eml{{\it emldetect}}
\def\sbl{SSB06}
\def\sbd{{\sc simbad}}

\def\lxlbol{$L_\mathrm{X}-L_\mathrm{bol}$}
\def\lx{$L_\mathrm{X}$}
\def\lbol{$L_\mathrm{bol}$}

\defcitealias{SBL98}{SBL98}
\defcitealias{SGR06}{Paper~I}

\title[An \xmm\ view of \ngc. II]{An \xmm\ view of the young open cluster NGC~6231\thanks{Based on observations collected with \xmm, an ESA science mission with instruments and contributions directly funded by ESA Member States and the USA (NASA). Table 2 and Figs. 2 to 5 are available from the CDS : http://cdsweb.u-strasbg.fr .} -- II. The OB star population}
\author[H.~Sana et al.~]
       {H. Sana$^{1,2}$\thanks{E-mail: hsana@eso.org}, G. Rauw$^1$\thanks{FNRS Research Associate (Belgium)},
 Y. Naz\'e$^1$\thanks{FNRS Postdoctoral Researcher (Belgium)},
 E. Gosset$^1$$\ddagger$\ and J.-M. Vreux$^1$\\
$^1$ Institut d'Astrophysique et de G\'eophysique, University of Li\`ege, All\'ee du 6 Ao\^ut 17, B\^at. B5c, B-4000 Li\`ege, Belgium\\
$^2$ European Southern Observatory, Alonso de Cordova 3107, Vitacura, Casilla 19001, Santiago 19, Chile}
\begin{document}

\date{Accepted 1988 December 15. Received 1988 December 14; in original form 1988 October 11}

\pagerange{\pageref{firstpage}--\pageref{lastpage}} \pubyear{2002}

\maketitle

\label{firstpage}

\begin{abstract}
 In this second paper of the series, we pursue the analysis of the 180~ks \xmm\ campaign towards the young open cluster NGC~6231 and we focus on its rich OB star population. We present a literature-based census of the OB stars in the field of view with more than one hundred objects, among which 30\% can be associated with an X-ray source. All the O-type stars are detected in the X-ray domain as soft and reasonably strong emitters. In the 0.5-10.0~keV band, their X-ray luminosities scale with their bolometric luminosities as $\log L_\mathrm{X} - \log L_\mathrm{bol}=-6.912\pm0.153$. Such a scaling law holds in the soft (0.5-1.0~keV) and intermediate (1.0-2.5~keV) bands but breaks down in the hard band. While the two colliding wind binaries in our sample clearly deviate from this scheme, the remaining O-type objects show a very limited dispersion (40\% or 20\% according to whether `cool' dwarfs are included or not),  much smaller than that obtained from previous studies. At our detection threshold and with our sample, the sole identified mechanism that produces significant modulations in the O star X-ray emission is related to wind interaction. We thus propose that the intrinsic X-ray emission of non-peculiar O-type stars can be considered as constant for a given star. In addition, the level of X-ray emission is accurately related to the star luminosity or, equivalently, to its wind properties.\\
Among B-type stars, the detection rate is only about 25\% in the sub-type range B0-B4 and remains mostly uniform throughout the different sub-populations while it drops significantly at later sub-types. The associated X-ray spectra are harder than those of O-type stars. Our analysis points towards the detected emission being associated with a physical (in a multiple system) PMS companion. However, we still observe a correlation between the bolometric luminosity of the B stars and the measured X-ray luminosity. The best fit power law  in the 0.5-10.0~keV band yields  $\log L_\mathrm{X} = 0.22(\pm0.06) \log L_\mathrm{bol}+22.8(\pm2.4)$. 
\end{abstract}


\section{Introduction} \label{sect: intro}

X-ray emission from early-type stars of spectral type O was, in December 1978, one of the earliest findings of the \einst\ satellite \citep{HBG79, SFG79}. 
It was soon realized that all O-type stars were X-ray emitters.
Most of them were characterized as soft (k$T<1$~keV) and reasonably strong ($10^{31} \la L_\mathrm{X} \la 10^{33}$~\ergs) sources. \citet{HBG79} already suggested that the X-ray luminosity was directly linked to the characteristic luminosity of the emitter, though at the time the authors proposed a scaling law between \lx\ and the visual luminosity. However as the stars of their sample had mostly the same colors, it is equivalent to assert that \lx\ scales with \lbol. Using various samples of stars observed with \einst, different authors \citep{LoW80, PGR81, CWS81, VCF81, SeC82} confirmed the so-called canonical relation $L_\mathrm{X} \approx 10^{-7}\ L_\mathrm{bol}$. Based on the \einst\ X-ray Observatory catalog of O-type stars \citep{CHS89}, \citet{SVH90} performed a  more comprehensive study of the relationship between the optical and X-ray properties of O-type stars. They confirmed the existence of a canonical \lxlbol\ relation, though with a rather large dispersion. They were unable to find any significant correlation with the rotation rate ($v \sin i$), the wind terminal velocity ($v_\infty$) or the mass-loss rate ($\dot{M}$), but observed a strong correlation with the wind momentum ($\dot{M} v_\infty)$ and with the wind luminosity ($0.5 \dot{M} v^2_\infty$). More recently, \citet{BSD97} investigated the properties of the bright OB-type stars detected in the \rosat\ all-sky survey \citep{BSC96} and found that the canonical relation extends down to spectral type B1-B1.5. They established the separation line between the O star relation and a less constrained relation for B stars to lie at a bolometric luminosity  $L_\mathrm{bol}\approx10^{38}$~\ergs.\\

Historically, two physically different models have been proposed to explain the hot O-type star X-ray emission. While no standard dynamo-driven surface magnetic field is theoretically expected for any star hotter than A7, \citet{CaO79} suggested a scaled-up version of a solar-type coronal emission model that yielded a roughly correct prediction for the X-ray flux but could not explain the softness of the  observed spectrum. Indeed in the coronal emission model, the X-rays are produced near the photosphere and are thus expected to suffer absorption by the overlying dense wind. Such an absorption is however not seen in the X-ray spectrum of hot stars. This  suggests that the X-ray emission is rather produced throughout a significant fraction of the wind volume, which has been a strong argument in favor of the embedded wind-shock model. 
In this second model, the X-ray emission is supposed to arise from shocks  occurring in the denser layers of the winds and that grow from small-scale instabilities of the line-driven winds. Since the phenomenological model of \citet{Luc82}, hydrodynamical simulations \citep[e.g.][]{OCR88,FPP97} have brought further support to this interpretation.\\

However, recent high resolution spectral observations of early-type stars are bringing the wind shock model to its limits. The unprecedented spectral resolution reached by the \xmm\ and \chandra\ observatories (corresponding to about 300~\kms\ in velocity space) now allows us to probe the widths and profiles of the X-ray emission lines seen in the spectra of hot stars. These contribute to put  constraints on the velocity (Doppler broadening) and location \citep[{\it fir} line ratio, ][]{PMD01} of the emitting plasma, yielding thus an unprecedented characterization of its localization in the expanding winds. \citet{KCO03} reported that the O-type supergiant $\zeta$  Puppis (O4~Ief) displays broad, blue-shifted and asymmetric line profiles that are generally consistent with the wind-shock model. However, observations of other early-type stars suggest  different pictures and hybrid magnetic wind models have been proposed. For example, the X-ray emission lines in the  spectrum of $\tau$~Scorpii (B0.2~V) are significantly narrower than expected from the standard wind-shock model \citep{CdMMF03}. These authors rather suggested magnetically confined wind shocks \citep{uDO02}, eventually coupled with the clump infall model \citep{HCB00}, as the prime origin for the observed X-ray emission. A similar model was also successfully applied to the magnetic rotator $\theta^1$~Ori~C \citep[O5.5V,][]{GOC05}. Finally, \citet{SCH03} have suggested that young massive stars could enter the main sequence carrying a significant residual magnetic field. Therefore, massive ZAMS stars could generate their X-ray luminosities via the standard model and magnetic confinement may provide an additional source of X-rays.\\

Compared to single stars of the same spectral type, close massive binaries are also known to display an extra X-ray emission \citep{ChG91} which is generally attributed to a wind-wind collision. Within the interaction region, the shocked gas is expected to be heated to temperatures of a few $10^7$~K and to generate a substantial amount of X-rays, which are thus produced in addition to the intrinsic emission by each of the components. This extra-emission can be further modulated  because e.g.\ of a variation of the optical depth along the line of sight towards the wind interaction zone due to the orbital motion. It could also reflect the changing properties of the shocks due, for example, to a variation of the distance between the two stars in an eccentric binary system. \\

While the earliest works based on \einst\ data suggested the canonical relation to extend through the B-type range and down to A5 stars \citep[e.g.][]{PGR81}, \citet{SGH85} showed that the latter law does not hold for A-type stars. Later works \citep{RGV85, CCM94} further suggested that the X-ray emission from  B stars of spectral type B2 or later was not following the same scheme as O-type stars. Because these B-type stars do not have the convective zones required to sustain a magnetic dynamo,  coronal emission is not expected. Their stellar winds are also much weaker than those of their hotter sisters, and are therefore not supposed to provide a sizable amount of X-rays. Indeed, \citet{BSD97} reported that, for stars of spectral type B2 or later, the detection rate drops below 10\%. Actually the intrinsic emission from B-type stars could, comparatively, be much lower than from O-type stars, with $\log \left( L_\mathrm{X}/L_\mathrm{bol} \right) \sim-8.5$ \citep{CCM97}. This suggests that most of the detected X-ray emission for B stars in distant clusters is actually associated with unresolved companions, either in a binary system or located by coincidence on the same line of sight. The question of the intrinsic B-type X-ray emission has however not yet received a satisfactory answer. One of the difficulties  is the intrinsic lack of homogeneity of the B-type population that contains different kinds of objects (\bcep, Be stars, shell stars, ...). An additional difficulty is their lower emission level (if any), which thus limits the number of detections on shorter duration exposures or in distant fields. It was hoped that the advent of  the `large' X-ray observatories, which combine improved sensitivity and spatial resolution, would help to solve this question. A scan of the recent literature \citep[e.g.][and references therein]{SHH03} indeed favours the `companion' scenario but provides by no mean a definitive answer to this question.\\

The young open cluster \ngc\ \citep[age $\sim$ 3 to 5~Myr,][]{BVF99} is considered as the core of the \sco\ association and contains a large OB star population. Located at about 1.6~kpc ($DM=11.07\pm0.04$, see discussion in \citealt{SGR06}, hereafter Paper~I), it offers an excellent opportunity to probe a homogeneous sample of early-type stars in terms of e.g.\ distance, age, reddening, environment and chemical composition. Our \xmm\ campaign has been described in  \citetalias{SGR06} and has revealed hundreds of point-like sources in the 15\arcmin\ radius field of view (FOV) of the satellite. In the present paper, we focus on the O- and B-type star population in the FOV. The analysis of the sources associated with optically faint counterparts is postponed to a forthcoming paper in this series (Sana et al. -- Paper III, in preparation).
Preliminary results of this work were presented in \citet{SNG06} but should be considered as supplanted by the present analysis.


This paper is organised as follows. The next section provides details of the data handling. Sect.~\ref{sect: ET} makes a census of the OB star population in the core of \ngc\ and identifies the early-type X-ray emitters. Sect.~\ref{sect: Xray} investigates the  X-ray properties of the detected O and B-type stars and the appropriate \lxlbol\ relations are derived in Sect.~\ref{ssect: lxlbol}. Sect.~\ref{sect: indiv} presents the properties of the individual early-type sources while  Sect.~\ref{sect: discuss} discusses the results of our study. Finally, Sect.~\ref{sect: ccl} summaries our main results.


\section{X-ray data handling} \label{sect: obs}

   \begin{figure}
     \centering
\vspace*{21.6cm}
     \caption{From top to bottom, combined \epicmos1, \mos2 and \pn\ images in the range 0.5-10.0\,keV. The background extraction regions adopted according to the different positions of the sources on the detectors have been overplotted. North is up, East to the left. A colour version of these figures is available in the electronic version of the paper.}
     \label{fig: bckg}
   \end{figure}

\begin{table}
\caption{ Positions and sizes ($r_\mathrm{extr.}$) of the  adopted background extraction regions in the FOV of the different \epic\ instruments. }
\label{tab: bkg}
\begin{tabular}{r r r r r}
\hline
Bkg. & Label & $\alpha$ (J2000.0) & $\delta$ (J2000.0) & $r_\mathrm{extr.}$  \\
\hline
\multicolumn{5}{c}{\mos1} \\
\hline
 CCD 1 & 1a  & 16$^\mathrm{h}$54$^\mathrm{m}$31\fs43 & $-$41\degr45\arcmin42\farcs2 & 20\arcsec \\ 
       & 1b  & 16$^\mathrm{h}$54$^\mathrm{m}$23\fs28 & $-$41\degr46\arcmin14\farcs7 & 20\arcsec \\ 
       & 1c  & 16$^\mathrm{h}$53$^\mathrm{m}$44\fs12 & $-$41\degr53\arcmin34\farcs6 & 25\arcsec \\ 
 CCD 2 & 2a  & 16$^\mathrm{h}$55$^\mathrm{m}$12\fs05 & $-$41\degr49\arcmin34\farcs0 & 50\arcsec \\ 
       & 2b  & 16$^\mathrm{h}$54$^\mathrm{m}$57\fs95 & $-$41\degr49\arcmin16\farcs7 & 50\arcsec \\ 
       & 2c  & 16$^\mathrm{h}$54$^\mathrm{m}$52\fs31 & $-$41\degr44\arcmin13\farcs0 & 35\arcsec \\ 
       & 2d  & 16$^\mathrm{h}$54$^\mathrm{m}$58\fs19 & $-$41\degr41\arcmin29\farcs2 & 45\arcsec \\ 
 CCD 3 & 3a  & 16$^\mathrm{h}$54$^\mathrm{m}$38\fs42 & $-$41\degr39\arcmin33\farcs3 & 30\arcsec \\ 
       & 3b  & 16$^\mathrm{h}$54$^\mathrm{m}$18\fs46 & $-$41\degr39\arcmin36\farcs0 & 40\arcsec \\ 
       & 3c  & 16$^\mathrm{h}$54$^\mathrm{m}$11\fs88 & $-$41\degr36\arcmin51\farcs0 & 30\arcsec \\ 
 CCD 4 & 4a  & 16$^\mathrm{h}$53$^\mathrm{m}$28\fs04 & $-$41\degr40\arcmin06\farcs8 & 35\arcsec \\ 
       & 4b  & 16$^\mathrm{h}$53$^\mathrm{m}$13\fs61 & $-$41\degr42\arcmin10\farcs2 & 45\arcsec \\ 
       & 4c  & 16$^\mathrm{h}$53$^\mathrm{m}$22\fs95 & $-$41\degr45\arcmin35\farcs4 & 45\arcsec \\ 
 CCD 5 & 5a  & 16$^\mathrm{h}$53$^\mathrm{m}$04\fs39 & $-$41\degr49\arcmin59\farcs9 & 50\arcsec \\ 
       & 5b  & 16$^\mathrm{h}$53$^\mathrm{m}$21\fs91 & $-$41\degr56\arcmin55\farcs4 & 60\arcsec \\ 
 CCD 6 & 6a  & 16$^\mathrm{h}$53$^\mathrm{m}$47\fs44 & $-$42\degr01\arcmin00\farcs9 & 45\arcsec \\ 
       & 6b  & 16$^\mathrm{h}$54$^\mathrm{m}$12\fs79 & $-$42\degr02\arcmin23\farcs5 & 45\arcsec \\ 
 CCD 7 & 7a  & 16$^\mathrm{h}$54$^\mathrm{m}$56\fs95 & $-$41\degr58\arcmin55\farcs4 & 50\arcsec \\ 
       & 7b  & 16$^\mathrm{h}$55$^\mathrm{m}$09\fs87 & $-$41\degr53\arcmin17\farcs6 & 60\arcsec \\ 
\hline
\multicolumn{5}{c}{\mos2} \\
\hline	    				       		       					   
 CCD 1 & 1a  & 16$^\mathrm{h}$54$^\mathrm{m}$31\fs43 & $-$41\degr45\arcmin42\farcs2 & 20\arcsec \\ 
       & 1b  & 16$^\mathrm{h}$54$^\mathrm{m}$23\fs28 & $-$41\degr46\arcmin14\farcs7 & 20\arcsec \\ 
       & 1c  & 16$^\mathrm{h}$53$^\mathrm{m}$44\fs12 & $-$41\degr53\arcmin34\farcs6 & 25\arcsec \\ 
 CCD 2 & 2a  & 16$^\mathrm{h}$54$^\mathrm{m}$12\fs79 & $-$42\degr02\arcmin38\farcs5 & 45\arcsec \\ 
       & 2b  & 16$^\mathrm{h}$54$^\mathrm{m}$55\fs16 & $-$41\degr58\arcmin50\farcs5 & 50\arcsec \\ 
 CCD 3 & 3a  & 16$^\mathrm{h}$55$^\mathrm{m}$09\fs87 & $-$41\degr53\arcmin17\farcs6 & 60\arcsec \\ 
       & 3b  & 16$^\mathrm{h}$55$^\mathrm{m}$12\fs05 & $-$41\degr49\arcmin35\farcs0 & 50\arcsec \\ 
       & 3c  & 16$^\mathrm{h}$54$^\mathrm{m}$57\fs95 & $-$41\degr49\arcmin16\farcs7 & 50\arcsec \\ 
 CCD 4 & 4a  & 16$^\mathrm{h}$54$^\mathrm{m}$52\fs53 & $-$41\degr43\arcmin55\farcs5 & 30\arcsec \\ 
       & 4b  & 16$^\mathrm{h}$54$^\mathrm{m}$57\fs29 & $-$41\degr41\arcmin14\farcs2 & 60\arcsec \\ 
       & 4c  & 16$^\mathrm{h}$54$^\mathrm{m}$38\fs53 & $-$41\degr39\arcmin29\farcs6 & 30\arcsec \\ 
       & 4d  & 16$^\mathrm{h}$54$^\mathrm{m}$24\fs48 & $-$41\degr39\arcmin38\farcs5 & 25\arcsec \\ 
 CCD 5 & 5a  & 16$^\mathrm{h}$53$^\mathrm{m}$49\fs69 & $-$41\degr37\arcmin34\farcs7 & 40\arcsec \\ 
       & 5b  & 16$^\mathrm{h}$53$^\mathrm{m}$28\fs04 & $-$41\degr40\arcmin06\farcs8 & 35\arcsec \\ 
 CCD 6 & 6a  & 16$^\mathrm{h}$53$^\mathrm{m}$22\fs95 & $-$41\degr45\arcmin35\farcs4 & 45\arcsec \\ 
       & 6b  & 16$^\mathrm{h}$53$^\mathrm{m}$01\fs20 & $-$41\degr49\arcmin59\farcs8 & 50\arcsec \\ 
 CCD 7 & 7a  & 16$^\mathrm{h}$53$^\mathrm{m}$21\fs91 & $-$41\degr56\arcmin55\farcs4 & 60\arcsec \\ 
       & 7b  & 16$^\mathrm{h}$53$^\mathrm{m}$47\fs44 & $-$42\degr01\arcmin00\farcs9 & 45\arcsec \\ 
\hline	
\end{tabular}
\end{table}
\begin{table}
\contcaption{ }			      
\begin{tabular}{r r r r r}
\hline
 Bkg. & Label & $\alpha$ (J2000.0) & $\delta$ (J2000.0) & $r_\mathrm{extr.}$ \\
\hline
\multicolumn{5}{c}{\pn} \\
\hline
   A   & 8b  & 16$^\mathrm{h}$53$^\mathrm{m}$51\fs25 & $-$41\degr37\arcmin04\farcs7 & 30\arcsec \\ 
       & 7b  & 16$^\mathrm{h}$54$^\mathrm{m}$11\fs43 & $-$41\degr36\arcmin48\farcs5 & 30\arcsec \\ 
   B   & 9b  & 16$^\mathrm{h}$53$^\mathrm{m}$22\fs96 & $-$41\degr39\arcmin56\farcs7 & 30\arcsec \\ 
       & 7a  & 16$^\mathrm{h}$54$^\mathrm{m}$20\fs24 & $-$41\degr39\arcmin36\farcs0 & 30\arcsec \\ 
       & 10b & 16$^\mathrm{h}$54$^\mathrm{m}$38\fs42 & $-$41\degr39\arcmin33\farcs3 & 30\arcsec \\ 
       & 11c & 16$^\mathrm{h}$54$^\mathrm{m}$58\fs19 & $-$41\degr41\arcmin29\farcs2 & 45\arcsec \\ 
   C   & 10a & 16$^\mathrm{h}$54$^\mathrm{m}$31\fs43 & $-$41\degr45\arcmin42\farcs2 & 20\arcsec \\ 
       & 11b & 16$^\mathrm{h}$54$^\mathrm{m}$54\fs54 & $-$41\degr44\arcmin28\farcs0 & 35\arcsec \\ 
       & 12b & 16$^\mathrm{h}$55$^\mathrm{m}$24\fs85 & $-$41\degr46\arcmin43\farcs3 & 60\arcsec \\ 
   D   & 9a  & 16$^\mathrm{h}$53$^\mathrm{m}$22\fs95 & $-$41\degr45\arcmin35\farcs4 & 45\arcsec \\ 
       & 8a  & 16$^\mathrm{h}$53$^\mathrm{m}$42\fs61 & $-$41\degr45\arcmin38\farcs3 & 30\arcsec \\ 
       & 11a & 16$^\mathrm{h}$55$^\mathrm{m}$04\fs20 & $-$41\degr48\arcmin31\farcs5 & 35\arcsec \\ 
       & 12a & 16$^\mathrm{h}$55$^\mathrm{m}$14\fs27 & $-$41\degr48\arcmin57\farcs4 & 30\arcsec \\ 
   E   & 6a  & 16$^\mathrm{h}$53$^\mathrm{m}$12\fs40 & $-$41\degr49\arcmin02\farcs6 & 30\arcsec \\ 
       & 6b  & 16$^\mathrm{h}$53$^\mathrm{m}$23\fs00 & $-$41\degr50\arcmin44\farcs2 & 30\arcsec \\ 
       & 5a  & 16$^\mathrm{h}$53$^\mathrm{m}$40\fs35 & $-$41\degr49\arcmin14\farcs5 & 25\arcsec \\ 
       & 2a  & 16$^\mathrm{h}$55$^\mathrm{m}$01\fs01 & $-$41\degr52\arcmin06\farcs6 & 35\arcsec \\ 
       & 3a  & 16$^\mathrm{h}$55$^\mathrm{m}$14\fs79 & $-$41\degr52\arcmin49\farcs9 & 60\arcsec \\ 
   F   & 6d  & 16$^\mathrm{h}$53$^\mathrm{m}$23\fs29 & $-$41\degr53\arcmin21\farcs2 & 40\arcsec \\ 
       & 5b  & 16$^\mathrm{h}$53$^\mathrm{m}$39\fs86 & $-$41\degr55\arcmin02\farcs0 & 20\arcsec \\ 
   G   & 6c  & 16$^\mathrm{h}$53$^\mathrm{m}$20\fs12 & $-$41\degr56\arcmin32\farcs9 & 40\arcsec \\ 
       & 4a  & 16$^\mathrm{h}$53$^\mathrm{m}$54\fs85 & $-$41\degr58\arcmin36\farcs0 & 30\arcsec \\ 
       & 1a  & 16$^\mathrm{h}$54$^\mathrm{m}$28\fs93 & $-$42\degr00\arcmin10\farcs9 & 45\arcsec \\ 
       & 2b  & 16$^\mathrm{h}$54$^\mathrm{m}$54\fs93 & $-$41\degr58\arcmin53\farcs0 & 50\arcsec \\ 
       & 3b  & 16$^\mathrm{h}$55$^\mathrm{m}$11\fs85 & $-$41\degr58\arcmin18\farcs7 & 45\arcsec \\ 
\hline
\end{tabular}
\end{table}


The present work relies on the X-ray source catalogue  and on the reduced X-ray data as obtained and described in Paper~I. Based on a census of the OB population in the FOV, we have now identified the early-type X-ray emitters in the FOV (see Sect.~\ref{sect: ET}). Using the SAS v6.0 software, we have then extracted the OB star X-ray spectra from the merged event lists, i.e. from the event lists built using the combination of the six pointings. For each object, we adopted a circular extraction region centered on the X-ray source positions listed in Paper~I and with a radius corresponding to half the distance to the nearest neighbouring X-ray source. Due to the crowded nature of the cluster core in the X-rays (see e.g. Fig.~1) and to the limited spatial resolution of the \epic\ detectors, the background could not be evaluated in the immediate vicinity of the stars, but had to be taken from the very few source free regions. We adopted several circular background regions spread throughout the FOV. For the two \epicmos\ instruments, as the background can be considered as uniform on a particular CCD, we associated, to each X-ray emitter, source-free background regions located on the same CCD detector. Regarding the \pn\ instrument, the background is rather considered to depend on the distance to the detector read-out nodes. We therefore selected source-free regions approximately situated at the same distance from the read-out nodes as the target of interest. The adopted background extraction regions for \mos\ and \pn\ instruments are displayed in Fig.~\ref{fig: bckg}. Their respective positions and sizes are given in Table \ref{tab: bkg}. We used the appropriate redistribution matrix files provided by the Science Operations Centre (SOC), according to the position of the considered source on the detectors. We built the corresponding ancillary response files using the arfgen command of the SAS software. The spectra were finally binned to have at least 10 counts per bin. The analysis of these spectra is presented in Sect.~\ref{sect: Xray}.

Finally, holding the source positions fixed at the values given in \citetalias{SGR06}, we ran the SAS task \eml\ on the individual pointings to obtain the background-subtracted, vignetting- and exposure-corrected count rates of the objects during each observation. Similarly, we used the extraction regions described above to build X-ray good time interval corrected and background subtracted light curves in the different energy bands considered in \citetalias{SGR06}. The latter curves  were computed for a series of time-bin sizes ranging from a few hundred seconds to 5~ks. These temporal information allow us to probe the variability of the early-type X-ray emitters on the different time-scales covered by the present data set (see Sect.~\ref{sect: indiv}).

\addtocounter{table}{+1}
\begin{table*}
\caption{List of the early-type X-ray emitters in the \xmm\ FOV. The first five columns give the source identifiers. Col. 6 indicates the separation $d_\mathrm{cc}$ between the associated optical and X-ray sources. Cols. 7 and 8 respectively provide the adopted instrument combination and the radius of the source extraction region. In Col. 7, the notations M1 and M2 respectively stand for \mos1 and \mos2. The adopted background regions are indicated in Cols. 9 to 11 for each \epic\ instrument (see Table \ref{tab: bkg}). The last three columns provide the bolometric luminosity ($L_\mathrm{bol}$), the ISM-corrected X-ray luminosity ($L_\mathrm{X}$) and the $L_\mathrm{X}/ L_\mathrm{bol}$ ratio respectively.  In the present table, the objects have been ordered by decreasing $L_\mathrm{bol}$. }
\label{tab: Xob}
\begin{tabular}{r r r r r r r r r r r r r r }
\hline
\vspace*{-3mm}\\
\multicolumn{5}{c}{Object IDs}   & $d_\mathrm{cc}$ & Instr. & $r_\mathrm{extr.}$ & \multicolumn{3}{c}{Background regions} & $\log L_\mathrm{bol}$ & $\log L_\mathrm{X}$ & $\log \frac{L_\mathrm{X}}{L_\mathrm{bol}}$ \\
  X\# & HD       & CPD$-41$\degr   & Se68 & SBL98 &(\arcsec)&    &(\arcsec)&{\sc M}1&{\sc M}2&\pn & (\ergs) & (\ergs)   \\
$[1]$ &    $[2]$ & $[3]$ & $[4]$& $[5]$ &$[6]$&  $[7]$  & $[8]$ &$[9]$&$[10]$&$[11]$& $[12]$ &  $[13]$ &  $[14]$   \\
\hline 
  279\hspace*{1.7mm}  &  152248  & 7728   &  291 &  856  & 0.8 &   \epic   & 19.0  &  1  & 1  & E  & 39.49 & 32.87  & $-$6.62  \\
  202\hspace*{1.7mm}  &  152234  & 7716   &  290 &  855  & 1.2 &   \mos    & 10.0  &  1  & 1  & -- & 39.45 & 32.42  & $-$7.03  \\
  216\hspace*{1.7mm}  &  152233  & 7718   &  306 &  858  & 1.3 &   \epic   & 14.0  &  1  & 1  & D  & 39.30 & 32.40  & $-$6.90  \\
  292\hspace*{1.7mm}  &  152249  & 7731   &  293 &  857  & 1.4 &   \epic   & 11.5  &  1  & 1  & E  & 39.22 & 32.30  & $-$6.92  \\
  291\hspace*{1.7mm}  &  152247  & 7732   &  321 &   --  & 0.5 &   \epic   & 23.0  &  3  & 5  & B  & 38.99 & 32.09  & $-$6.90  \\
  422\hspace*{1.7mm}  &  326331  & 7744   &  338 &  571  & 0.7 &   \mos    & 12.0  &  1  & 1  & -- & 38.92 & 31.96  & $-$6.96  \\
  468\hspace*{1.7mm}  &  152314  & 7749   &  161 &  615  & 0.8 &   \epic   & 15.5  &  1  & 1  & D  & 38.90 & 31.80  & $-$7.10  \\
  185\hspace*{1.7mm}  &  152218  & 7713   &    2 &  853  & 1.1 &   \mos    & 20.0  &  3  & 5  & -- & 38.77 & 31.87  & $-$6.90  \\
  149\hspace*{1.7mm}  &  152219  & 7707   &  254 &  234  & 0.9 &   \epic   & 08.0  &  1  & 1  & F  & 38.68 & 31.80  & $-$6.88  \\
  306\hspace*{1.7mm}  &    --    & 7733   &  297 &  862  & 1.1 &   \epic   & 11.5  &  1  & 1  & E  & 38.62 & 31.83  & $-$6.79  \\
  372\hspace*{1.7mm}  &    --    & 7742   &  224 &  505  & 1.9 &   \mos    & 13.0  &  1  & 1  & -- & 38.48 & 31.94  & $-$6.54  \\
  126\hspace*{1.7mm}  &  152200  & 7702   &  266 &  206  & 0.7 &   \epic   & 10.0  &  1  & 1  & E  & 38.44 & 31.29  & $-$7.15  \\
    6\hspace*{1.7mm}  &  152076  & 7684   &   -- &   --  & 1.2 &   \mos    & 37.5  &  4  & 6  & -- & 38.44 & 31.04  & $-$7.40  \\
  566$^\mathrm{a}$    &    --    & 7760AB &  810 &   --  & 1.6 &   \pn     & 17.0  &  -- & -- & G  & 38.44 & 30.19  & $-$8.25  \\
  313\hspace*{1.7mm}  &  326329  & 7735   &  292 &  434  & 0.5 &   \epic   & 07.5  &  1  & 1  & E  & 38.33 & 31.78  & $-$6.55  \\
  251\hspace*{1.7mm}  &    --    & 7721p  &  309 &  350  & 1.0 &   \epic   & 10.0  &  1  & 1  & E  & 38.31 & 31.44  & $-$6.87  \\
  448\hspace*{1.7mm}  &    --    & 7746   &  769 &   --  & 1.7 &   \mos    & 07.0  &  3  & 4  & -- & 38.06 & 30.79  & $-$7.27  \\
  139\hspace*{1.7mm}  &    --    & 7706   &  253 &  226  & 1.2 &   \mos    & 17.0  &  1  & 1  & -- & 37.74 & 31.17  & $-$6.57  \\
  521\hspace*{1.7mm}  &    --    & 7755   &  112 &  684  & 1.7 &   \mos    & 13.0  &  7  & 3  & -- & 37.59 & 31.08  & $-$6.51  \\
  167\hspace*{1.7mm}  &  326320  & 7710   &  745 &   --  & 0.5 &   \epic   & 16.0  &  3  & 5  & B  & 37.57 & 30.60  & $-$6.97  \\
  203\hspace*{1.7mm}  &    --    & 7715   &  261 &  303  & 1.1 &   \epic   & 11.0  &  1  & 1  & E  & 37.36 & 30.87  & $-$6.49  \\
  337$^\mathrm{a}$    &    --    & 7737   &  294 &  461  & 0.7 &   \epic   & 07.5  &  1  & 1  & E  & 37.19 & 31.20  & $-$5.99  \\
  213\hspace*{1.7mm}  &   --     &  --    &  259 &  317  & 1.5 &   \epic   & 11.0  &  1  & 1  & E  & 37.15 & 30.97  & $-$6.18  \\
  282\hspace*{1.7mm}  &    --    &  --    &  209 &  394  & 1.2 &   \epic   & 13.0  &  1  & 1  & D  & 37.14 & 31.12  & $-$6.02  \\
   22\hspace*{1.7mm}  &  326343  & 7688   &  616 &   19  & 1.4 &   \epic   & 08.0  &  5  & 7  & F  & 37.13 & 30.26  & $-$6.87  \\
  440\hspace*{1.7mm}  &    --    &  --    &  160 &  593  & 1.2 &   \epic   & 08.0  &  1  & 1  & D  & 36.48 & 30.67  & $-$5.81  \\
  559\hspace*{1.7mm}  &    --    &  --    &  142 &  748  & 0.8 &{\sc M}2+\pn& 15.0 &  -- & 3  & E  & 36.39 & 30.66  & $-$5.73  \\
   78\hspace*{1.7mm}  &    --    &  --    &   41 &  149  & 1.3 &   \epic   & 15.0  &  1  & 1  & E  & 36.28 & 30.85  & $-$5.43  \\
  557\hspace*{1.7mm}  &    --    &  --    &  127 &  746  & 0.9 &   \epic   & 11.0  &  7  & 3  & E  & 36.08 & 30.73  & $-$5.35  \\
  121\hspace*{1.7mm}  &    --    &  --    &  265 &  200  & 1.2 &   \epic   & 09.0  &  1  & 1  & E  & 35.64 & 30.81  & $-$4.83  \\
\hline									
\end{tabular}\vspace*{1mm} \\
\begin{flushleft}
$^a$ See individual comments in Sect.~\ref{ssect: Bindiv}\\
\end{flushleft}
\end{table*}


\section{Early-type stars in \ngc} \label{sect: ET}

We first take a census of the OB star population in the observed FOV. Relying on a large review of the literature (see references in \citetalias{SGR06}), on the Catalog of Galactic OB Stars \citep{Reed03}, and on the WEBDA\footnote{http://obswww.unige.ch/webda/} and SIMBAD\footnote{http://simbad.u-strasbg.fr/Simbad/} databases, we selected all stars for which at least one of the previous references quoted a spectral type O or B, regardless of their sub-spectral type, their luminosity classification or their single/multiple status. This resulted in 106 objects among which 91 B stars and 15 O stars. These are listed in Table 2 available from the CDS (http://cdsweb.u-strasbg.fr). \citet{Hou78} quoted HD~152437 as a B9 star. However, we recently revised its classification to A0III \citep{phd} and estimated its distance to about 500~pc. This object is thus not included in the present census.
 Together with various cross-identifications (Cols. 1 to 7), Table 2  provides the adopted positions (Col. 8) and the $V$ and $B-V$ magnitudes (Cols. 9 and 10) obtained from the \sbl\ catalogue (Sung et al. 2006, in preparation; see also \citetalias{SGR06} for a brief description). 
The visual ($M_\mathrm{V}$) and bolometric ($M_\mathrm{bol}$) absolute magnitudes (Cols. 11 and 12) were computed using a distance modulus $DM=11.07$, the intrinsic colours and bolometric correction scale of \citet{SK82} and a reddening law $A_\mathrm{V}=R\times E(B-V)$ with $R=3.3$ \citep{SBL98}. $A_\mathrm{V}$ was computed for each object to account for the differential reddening existing across the cluster \citep{SBL98}. The adopted spectral types and the corresponding reference are quoted  in Cols. 13 and 14. For binaries, the quoted visual and bolometric magnitudes are given for the system as a whole. We used the respective visual brightness ratio of the system components and derived a reddening correction that accounts for the component spectral types and for their relative contribution to the total light from the system. Doing so, we deduced the $M_\mathrm{V}$ for the system. We then derived the individual visual and bolometric magnitudes. Finally, we combined the latter values to obtain the total bolometric magnitude for the system. The last column (Col. 15) of Table 2 provides miscellaneous notes about the objects, together with the related references. Compared to the previous results presented in \citet{SNG06}, some spectral types have been revised thanks to recent FEROS observations \citep{phd}. The distribution of the selected stars among the different spectral sub-types and luminosity classes given in \citet{SNG06} still provides a good overview of the early-type population in \ngc.\\

As a next step, we cross-correlated the X-ray catalogue of \citetalias{SGR06} with the obtained list of OB stars in the FOV. We first built the distribution of the number of associated OB counterparts as a function of the cross-correlation radius. The curve reveals 30 associations within a cross-correlation distance ($d_\mathrm{cc}$) of 2\farcs5. Six additional sources are found with $d_\mathrm{cc}$ between 3\farcs0 and 3\farcs6. The remaining associations require $d_\mathrm{cc}>4\farcs8$.
For consistency with the results of \citetalias{SGR06}, we adopted a cut-off radius of 2\farcs5, thus resulting in 30 OB X-ray emitters (see Table~\ref{tab: Xob}). Except in one case, the associated early-type stars are the only optical counterpart found in the \sbl\ catalogue within the adopted cut-off radius. Considering that faint optical sources located close to an early-type star (which is intrinsically bright) are difficult to detect and might indeed be missing in the SSB06 catalogue, one can however estimate the probability to spuriously associate an OB star with the X-ray emission produced by a lower mass star located by chance on the same line of sight. Using the \citetalias{SGR06} results and considering only stars with $V<19$, one obtains that this probability remains below 0.03. As a consequence, the early-type stars listed in Table~\ref{tab: Xob} are  most probably physically related to the observed X-ray sources.\\
\begin{table*}
\centering
\caption{Best-fit single temperature  models of the form: {\tt wabs$_\mathrm{ISM}$ * wabs * mekal}. The first and second columns give the X-ray and optical identifiers. Col. 3  provides the adopted interstellar column ($N_\mathrm{H}^\mathrm{ISM}$, in $10^{22}$\,cm$^{-2}$). The next three columns (Cols. 4 to 6) list the best-fit absorbing column ($N_\mathrm{H}$, in $10^{22}$\,cm$^{-2}$), temperature (k$T$, in keV) and normalisation factor (in cm$^{-5}$, $norm =\frac{10^{-14}}{4\pi d^2}\int N_\mathrm{e} N_\mathrm{H} dV$ with $d$, the distance to the source -- in cm --, $N_\mathrm{e}$ and $N_\mathrm{H}$, the electron and hydrogen number densities -- in cm$^{-3}$). The quoted upper and lower values indicate the limits of the 90\% confidence interval (see text). The reduced chi-square and the number of degrees of freedom (dof) of the fit are given in Col. 7.  Cols. 8 and 9 provide the observed ($f_\mathrm{X}$) and the ISM absorption-corrected  ($f_\mathrm{X}^\mathrm{corr.}$) fluxes (in $10^{-14}$\,\ergscm) in the range 0.5-10.0\,keV. The last column provides the sum of the numbers of source counts in the fitted spectra.  The horizontal line separates the O-type objects (upper part of the table) from the B-type stars (bottom part of the table). }
\label{tab: Xspec_1T}
\begin{tabular}{r r r r r r r r r r}
\hline
 \vspace*{-3mm}\\
X \# &    ID  & $N_\mathrm{H}^\mathrm{ISM}$ & $N_\mathrm{H}$ & k$T$ & $norm$ & \chin\ (dof) & $f_\mathrm{X}$ &  $f_\mathrm{X}^\mathrm{corr.}$ & Net cts \\
$[1]$ & $[2]$ & $[3]$ & $[4]$ & $[5]$ & $[6]$ & $[7]$ & $[8]$ & $[9]$ & $[10]$  \\
\hline
    6 &   152076  & 0.29 & $<0.12$              & $0.28^{0.36}_{0.23}$ & $1.85^{4.20}_{1.20}\times 10^{-5}$ & 1.08 (92) & 0.92 & 3.42 & 1095\\
\hline
   22 &   326343  & 0.36 & $0.43^{0.87}_{0.09}$ & $0.56^{0.84}_{0.28}$ & $ < 4.73 \times 10^{-5}          $ & 1.08 (13) & 0.25 & 0.57 &  170\vspace*{1mm}\\

  167 &   326320  & 0.24 & $0.61^{0.73}_{0.47}$ & $0.61^{0.82}_{0.50}$ & $2.50^{3.85}_{1.39}\times 10^{-5}$ & 0.94 (70) & 0.79 & 1.24 &  762\vspace*{1mm}\\

  448 &     7746  & 0.30 & $0.48^{0.67}_{0.27}$ & $0.58^{0.75}_{0.43}$ & $2.99^{6.00}_{0.00}\times 10^{-5}$ & 1.12 (11) & 1.01 & 1.92 &  144\vspace*{1mm}\\

  566 &     7760AB  & 0.42 & $0.79^{1.69}_{0.00}$ & $0.28^{1.74}_{0.08}$ & $ < 1.71                         $ & 0.76 (23) & 0.18 & 0.48 &  265\\
\hline
\end{tabular}
\end{table*}

 After a visual inspection of the X-ray images at the position of the undetected OB stars, we further added two objects to the list of early-type X-ray emitters. Indeed it clearly appears that  the related locations on the \epic\ images  have been corrupted because of the crowdedness of the field in these particular regions.  The spectra of these sources could of course not be securely extracted because of the probable contamination by very close X-ray neighbours. These two sources are thus not included in the discussions of Sect.~\ref{sect: Xray}; still they are very probable X-ray emitters. These two additional sources are identified with a cross (+) appended to the ID number of the X-ray source (Col.~6) in Table 2. For these two objects, we emphasize that the corresponding X-ray source in the catalogue of \citetalias{SGR06} is probably constituted of different components that the psf (point spread function) fit has not been able to separate. Finally, we found residuals larger than expected for source-free regions at the locations of seven B-type stars. These might be X-ray sources below the detection threshold of our catalogue (see details in \citetalias{SGR06}). For the sake of completeness, we also identified these stars in Table 2. \\

\begin{table*}
\centering
\caption{Same as Table \ref{tab: Xspec_1T} for two-temperature  models of the form: {\tt wabs$_\mathrm{ISM}$ * (wabs$_1$ * mekal$_1$ + wabs$_2$ * mekal$_2$)}. $norm_1$ and $norm_2$ are expressed in $10^{-5}$~cm$^{-5}$.}
\label{tab: Xspec_2T}
\begin{tabular}{r r r r r r r r r r r r r}
\hline
 \vspace*{-3mm}\\
X \# &    ID  & $N_\mathrm{H}^\mathrm{ISM}$ & $N_\mathrm{H,1}$ & k$T_1$ & $norm_1$ & $N_\mathrm{H,2}$ & k$T_2$ & $norm_2$ & \chin\ (dof) & $f_\mathrm{X}$ &  $f_\mathrm{X}^\mathrm{corr.}$  & Net cts\\
$[1]$ & $[2]$ & $[3]$ & $[4]$ & $[5]$ & $[6]$ & $[7]$ & $[8]$ & $[9]$ & $[10]$ & $[11]$ & $[12]$ & $[13]$  \\
\hline
  126 &   152200 & 0.23  & $     <0.06        $ & $0.36^{0.40}_{0.32}$ & $1.94^{2.41}_{1.58}$ 
                         & $0.23^{0.38}_{0.10}$ & $0.71^{0.78}_{0.65}$ & $1.51^{1.93}_{1.11}$ & 1.35 (157)& 2.80 &  6.13 &  2079\vspace*{1mm}\\

  149 &   152219 & 0.26  &   n.                 & $0.26^{0.29}_{0.22}$ & $7.47^{8.19}_{6.49}$ 
                         & $0.22^{0.35}_{0.11}$ & $0.67^{0.72}_{0.63}$ & $5.01^{5.38}_{4.17}$ & 1.09 (268)& 6.90 & 19.50 &  3674\vspace*{1mm}\\

  185 &   152218 & 0.26  & $0.01^{0.17}_{0.00}$ & $0.31^{0.34}_{0.27}$ & $8.82^{24.70}_{7.37}$ 
                         & $0.40^{0.52}_{0.30}$ & $0.71^{0.76}_{0.65}$ & $7.99^{9.63}_{6.65}$ & 0.95 (159)& 9.49 & 23.24 &  3286\vspace*{1mm}\\

  202 &   152234 & 0.22  & $0.22^{0.29}_{0.12}$ & $0.27^{0.29}_{0.25}$ & $105.^{166.}_{057.}$ 
                         & $0.67^{0.81}_{0.58}$ & $0.76^{0.80}_{0.72}$ & $35.6^{39.8}_{31.2}$ & 1.47 (234)& 39.74& 82.23 & 12255\vspace*{1mm}\\

  251 &     7721p& 0.24  & $0.41^{0.48}_{0.31}$ & $0.24^{0.28}_{0.22}$ & $34.5^{54.2}_{30.2}$ 
                         & $1.38^{1.89}_{0.46}$ & $1.33^{1.90}_{1.03}$ & $2.48^{4.50}_{1.44}$ & 0.92 (204)& 4.07 & 8.56  &  2797\vspace*{1mm}\\

  291 &   152247 & 0.28  & $<0.06             $ & $0.25^{0.28}_{0.21}$ & $1.49^{36.3}_{13.0}$  
                         & $0.12^{0.24}_{0.03}$ & $0.63^{0.69}_{0.59}$ & $7.41^{8.67}_{5.89}$ & 1.26 (351)& 12.20& 37.90 &  6932\vspace*{1mm}\\

  292 &   152249 & 0.27  & $0.25^{0.30}_{0.21}$ & $0.23^{0.24}_{0.22}$ & $145.^{202.}_{109.}$  
                         & $1.18^{1.70}_{0.86}$ & $0.81^{0.93}_{0.73}$ & $12.3^{16.6}_{ 9.4}$ & 1.40 (389)& 21.39& 61.85 & 17380\vspace*{1mm}\\

  372 &     7742 & 0.28  &   n.                 & $0.59^{0.62}_{0.50}$ & $ 7.4^{ 8.0}_{ 6.2}$ 
                         & $0.73^{0.86}_{0.62}$ & $1.05^{1.26}_{0.97}$ & $13.9^{15.9}_{12.9}$ & 1.09 (209)& 14.0 & 27.2  &  4790\vspace*{1mm}\\

  422 &   326331 & 0.28  & $0.29^{0.42}_{0.19}$ & $0.23^{0.25}_{0.20}$ & $77.8^{173.}_{39.7}$ 
                         & $1.23^{1.85}_{0.88}$ & $0.88^{1.02}_{0.76}$ & $10.6^{15.8}_{ 7.9}$ & 1.19 (157)& 10.70& 28.74 &  3325\vspace*{0mm}\\
\hline
   78 &   Se~41  & 0.21  & $0.65^{0.79}_{0.38}$ & $0.21^{0.29}_{0.19}$ & $21.6^{66.4}_{ 8.1}$ 
                         & $0.76^{1.86}_{0.06}$ & $1.66^{2.56}_{1.20}$ & $1.18^{1.86}_{0.77}$ & 1.11 (74) & 1.31 & 2.02  &  1198\vspace*{1mm}\\

  121 &   Se~265 & 0.26  & $0.52^{0.69}_{0.26}$ & $0.19^{0.23}_{0.14}$ & $18.7^{76.7}_{ 3.5}$ 
                         & $0.49^{1.03}_{0.24}$ & $1.83^{2.47}_{1.43}$ & $1.37^{1.88}_{0.77}$ & 1.37 (105)& 1.34 & 2.22  &   789\vspace*{1mm}\\

  139 &     7706 & 0.25  & $0.62^{0.84}_{0.37}$ & $0.26^{0.38}_{0.19}$ & $13.2^{69.7}_{ 2.8}$ 
                         & $            <0.51 $ & $2.56^{3.25}_{1.82}$ & $1.98^{2.84}_{1.64}$ & 0.99 (92) & 3.02 & 4.62  &  1183\vspace*{1mm}\\

  203 &    7715  & 0.25  & $0.58^{0.79}_{0.29}$ & $0.23^{0.27}_{0.18}$ & $12.8^{63.2}_{ 5.3}$ 
                         & $1.00^{2.15}_{0.60}$ & $1.32^{1.82}_{1.08}$ & $1.87^{2.89}_{1.29}$ & 1.29 (94) & 1.42 & 2.32  &  1078\vspace*{1mm}\\

  213 &  Se~259  & 0.26  & $0.75^{0.87}_{0.61}$ & $0.15^{0.16}_{0.14}$ & $185.^{383.}_{121.}$ 
                         & $ < 1.77           $ & $2.00^{3.33}_{1.26}$ & $1.60^{3.03}_{1.06}$ & 1.10 (100)& 1.75 & 2.88  &  1151\vspace*{1mm}\\

  282 &  Se~209  & 0.24  & $0.76^{0.83}_{0.64}$ & $0.16^{0.17}_{0.14}$ & $169.^{280.}_{ 72.}$  
                         & $0.71^{1.05}_{0.32}$ & $1.65^{2.20}_{1.45}$ & $3.04^{3.81}_{2.30}$ & 1.02 (168)& 2.66 & 4.06  &  1965\vspace*{1mm}\\

  337 &     7737 & 0.26  & $0.57^{0.70}_{0.39}$ & $0.24^{0.37}_{0.22}$ & $21.7^{36.9}_{10.5}$ 
                         & $0.52^{0.92}_{0.00}$ & $1.76^{2.55}_{1.50}$ & $2.93^{3.59}_{1.96}$ & 0.91 (122)& 3.09 &  4.95 &  1418\vspace*{1mm}\\

  440 &  Se~160  & 0.29  & $0.85^{1.13}_{0.00}$ & $0.23^{0.42}_{0.08}$ & $16.1^{79.8}_{ 0.9}$ 
                         & $0.18^{6.06}_{0.00}$ & $3.10^{13.93}_{1.84}$& $0.57^{1.41}_{0.33}$ & 0.82 (35) & 0.96 & 1.45  &   432\vspace*{1mm}\\

  521 &    7755  & 0.27  &    n.                & $0.63^{0.75}_{0.47}$ & $0.54^{0.70}_{0.40}$ 
                         & $ < 0.39$            & $2.28^{2.86}_{1.74}$ & $1.66^{2.08}_{1.39}$ & 1.02 (43) & 2.31 & 3.72  &   523\vspace*{1mm}\\

  557 &  Se~127  & 0.33  & $0.19^{0.81}_{0.00}$ & $0.27^{0.69}_{0.19}$ & $1.05^{13.44}_{0.33}$
                         & $0.92^{1.87}_{0.47}$ & $1.11^{1.56}_{0.90}$ & $1.92^{3.49}_{1.26}$ & 0.88 (42) & 0.91 & 1.66  &   492\vspace*{1mm}\\

  559 &  Se~142  & 0.29  & $ < 0.85           $ & $0.68^{0.88}_{0.38}$ & $0.22^{1.30}_{0.12}$
                         & $ < 2.32           $ & $3.16^\mathrm{ n. }_{1.70}$ & $0.62^{1.30}_{0.18}$ & 0.87 (43) & 0.93 & 1.42 &   506\\
\hline
\end{tabular}
\end{table*}

As a first result, the 15 O-type objects in the FOV are all seen in the X-rays while only $\sim$25\% (15 out of 63) of the early  B stars (sp.\ type B0--B4) are detected. This percentage remains almost constant within the subtypes B0 to B4 but reaches $\sim45$\% among the giants and supergiants (4 out of 9). Except for the respectively known (\cpd7737) and suspected (Se~209) slowly pulsating B-type stars (SPB) in the FOV, that are both detected in the X-ray domain, no particular correlation is found for the different sub-populations. Three out of the six known \bcep\ are detected (alternatively, three out of the eight known or suspected \bcep\ are detected). Five out of the 16 known binaries are also detected (alternatively, six out of the 26 known or suspected binaries are also detected). For the later B-type stars (subtypes B5 and later), only two objects out of 28 are detected, which constitutes a much smaller detection rate. 


\section{X-ray properties }\label{sect: Xray}

\subsection{X-ray spectrum\label{ssect: Xspec} }

Using the procedure described in Sect.~\ref{sect: obs}, we extracted the merged X-ray spectra of each OB  emitters. We selected the deepest instrument combination (see Table \ref{tab: Xob}), requiring that the source is not located too close to a gap or a detector edge. The radius of the source extraction region and the identification of the associated background regions in the different instruments are given in Table \ref{tab: Xob}. The spectra are relatively soft and peak at an energy well below $1.5$~keV. These X-ray spectra are presented in Figs. 2 to 5, also available from the CDS. Beyond the different emission levels, the O- and B-type spectra show clearly different characteristics. 
\addtocounter{figure}{+4}

To quantify the physical properties of the X-ray emission, we adjusted a series of optically thin thermal plasma \mek\ models \citep{MGvdO85, Ka92} to the obtained spectra. For the purpose of scientific analysis, we limited the considered energy range to the 0.5-10.0\,keV band and we simultaneously fitted the different \epic\ spectra using the \xspec\ software v.11.2.0\footnote{http://xspec.gsfc.nasa.gov/} \citep{arn96}. Because of the expected presence of local absorption (e.g.~due to the winds), we requested, in the spectral fits, a total equivalent column density of  neutral hydrogen larger or equal to the interstellar (ISM) column density $N_{\rm H}^{\rm ISM}$. The latter was estimated from the color excess using the relation of \citet{BSD78}: $N_\mathrm{H}= E(B-V)\times 5.8\times10^{21}$~cm$^{-2}$. Fits with a common local absorption column for multi-temperature models were, for various cases (e.g.~\hd234, \hd248, \hd249, ...), unable to  reproduce the X-ray spectra. We thus adopted independent local absorbing columns for each \mek\ component in the model. As no strong deviation from the solar metallicity is seen in the optical spectra, we kept the chemical composition fixed at the solar values. We successively tested one-, two- and three-temperature absorbed \mek\ models.  According to the quality  of the observed spectrum, we finally adopted the model with the lowest number of parameters that still adequately reproduces the data. Good spectral fits are generally obtained using two-temperature (2-T) \mek\ models with independent local absorbing columns for each component. However, the brightest sources may require an additional, higher temperature component (3-T models) while the spectra of a few very faint sources are satisfactorily described by a single temperature (1-T) component.  The observed emission is generally soft (k$T \le 2.0$\,keV) and slightly absorbed ($N_\mathrm{H}\le 10^{22}$\,cm$^{-2})$. Best fit results are given in Tables \ref{tab: Xspec_1T} to \ref{tab: Xspec_3T} together with the observed and ISM-absorption corrected fluxes in the range 0.5-10.0~keV. Tables \ref{tab: Xspec_1T} to \ref{tab: Xspec_3T} also provide the 90\% confidence intervals on the best fit parameters. These have been computed using the \xspec\ command {\it error}. In this procedure, each parameter is varied until the value of the $\chi^2$, minimized by allowing all the other non-frozen parameters to vary, is equal to the best fit value increased by a value $\Delta \chi^2=2.706$, equivalent to the 90\% confidence region for the considered parameter.

Direct comparison of the spectral properties obtained with models composed of a different number of \mek\ components is often not an easy step. For this reason, the different result tables were organised according to the number of \mek\ components used in the model.  Table \ref{tab: Xspec_2T} probably provides the best illustration of the clearly distinct X-ray spectral characteristics of the O- and B-type stars. 
With the  temperatures of the two \mek\ components mostly below 1~keV, the X-ray emission from the O-type stars is particularly soft. The first temperature is about $0.3$~keV, except for \cpd7742. However, \citet{SAR05} showed that the intrinsic X-ray emission from this system is generally contaminated by an extra-component resulting from an ongoing wind interaction process, except when the interaction region is occulted. At these particular phases at which we probably only detect the intrinsic emission from the O-type primary, it is interesting to note that the temperature of the `cool' component drops to about 0.35~keV \citep{SAR05}, thus quite close to the average `cool' temperature of the other O-type objects.
\begin{table*}
\centering
\caption{Same as Table \ref{tab: Xspec_1T} for three-temperature  models of the form: {\tt wabs$_\mathrm{ISM}$ * $\sum_i$ (wabs$_i$*mekal$_i$)}, with $i=1, 3$. The fourth column gives the number $i$ of the considered component in the model.}
\label{tab: Xspec_3T}
\begin{tabular}{r r r r r r r r r r r}
\hline
 \vspace*{-3mm}\\
X \# & Object ID. & $N_\mathrm{H}^\mathrm{ISM}$ & $i$ &  $N_{\mathrm{H},\,i}$ & k$T_i$ & $norm_i$ & \chin\ (dof) & $f_\mathrm{X}$ &  $f_\mathrm{X}^\mathrm{corr.}$ & Net cts \\
$[1]$ & $[2]$ & $[3]$ & $[4]$ & $[5]$ & $[6]$ & $[7]$ & $[8]$ & $[9]$ & $[10]$ & $[11]$  \\

\hline

  216 &   152233 & 0.25 & 1 & $0.53^{0.60}_{0.48}$ & $0.15^{0.16}_{0.14}$ & $1.47^{2.46}_{1.10}\times 10^{-2}$ &           &      &       &      \vspace*{1mm}\\
      &          &      & 2 & $     <0.05        $ & $0.48^{0.51}_{0.44}$ & $1.17^{1.43}_{1.07}\times 10^{-4}$ &           &      &       &      \vspace*{1mm}\\
      &          &      & 3 & $0.89^{1.10}_{0.78}$ & $0.64^{0.69}_{0.61}$ & $4.10^{4.72}_{3.53}\times 10^{-4}$ & 1.74 (544)& 33.16& 78.20 & 26522\vspace*{1mm}\\

  279 &   152248 & 0.26 & 1 &    n.                & $0.29^{0.39}_{0.19}$ & $7.41^{7.86}_{6.96}\times 10^{-4}$ &           &      &       &      \vspace*{1mm}\\
      &          &      & 2 & $0.44^{0.47}_{0.41}$ & $0.64^{0.79}_{0.49}$ & $1.20^{1.33}_{1.07}\times 10^{-3}$ &           &      &       &      \vspace*{1mm}\\
      &          &      & 3 &    n.                & $4.75^{1.50}_{8.00}$ & $2.38^{4.31}_{0.46}\times 10^{-5}$ & 2.40 (783)& 98.1 & 232.  & 91327\vspace*{1mm}\\

 		  				
%
  306 &    7733  & 0.26 & 1 & n.                   & $0.26^{0.28}_{0.25}$ & $0.99^{1.03}_{0.93}\times 10^{-4}$ &           &      &       &      \vspace*{1mm}\\ 
      &          &      & 2 & $0.45^{0.56}_{0.35}$ & $0.65^{0.70}_{0.61}$ & $6.42^{7.20}_{5.13}\times 10^{-5}$ &           &      &       &      \vspace*{1mm}\\ 
      &          &      & 3 & $6.00^{15.7}_{1.14}$ & $2.36^{9.25}_{1.18}$ & $0.16^{1.02}_{0.00}\times 10^{-4}$ & 1.29 (311)& 8.12 & 22.4  &  5955\vspace*{1mm}\\

  313 &   326329 & 0.22 & 1 & $ < 0.34 $           & $0.30^{0.34}_{0.19}$ & $3.51^{28.64}_{2.69}\times10^{-5}$ &           &      &       &      \vspace*{1mm}\\ 
      &          &      & 2 & $0.33^{0.43}_{0.20}$ & $0.66^{0.75}_{0.60}$ & $7.78^{10.40}_{5.07}\times10^{-5}$ &           &      &       &      \vspace*{1mm}\\ 
      &          &      & 3 & $ < 0.19 $           & $2.62^{3.25}_{2.21}$ & $2.93^{3.54}_{2.32}\times 10^{-5}$ & 1.24 (323)& 9.96 & 18.58 &  4927\vspace*{1mm}\\

  468 &   152314 & 0.30 & 1 & $0.11^{0.17}_{0.00}$ & $0.24^{0.26}_{0.19}$ & $1.64^{5.46}_{0.74}\times 10^{-4}$ &           &      &       &      \vspace*{1mm}\\ 
      &          &      & 2 & $0.21^{0.39}_{0.06}$ & $0.68^{0.78}_{0.60}$ & $3.11^{4.59}_{1.81}\times 10^{-5}$ &           &      &       &      \vspace*{1mm}\\ 
      &          &      & 3 & $    < 4.89        $ & $4.70^{63.0}_{2.18}$ & $5.65^{10.60}_{3.48}\times10^{-6}$ & 1.07 (296)& 6.53 & 19.85 &  4971\\
\hline
\end{tabular}
\end{table*}
 
   \begin{figure}
     \centering
     \includegraphics[width=\columnwidth]{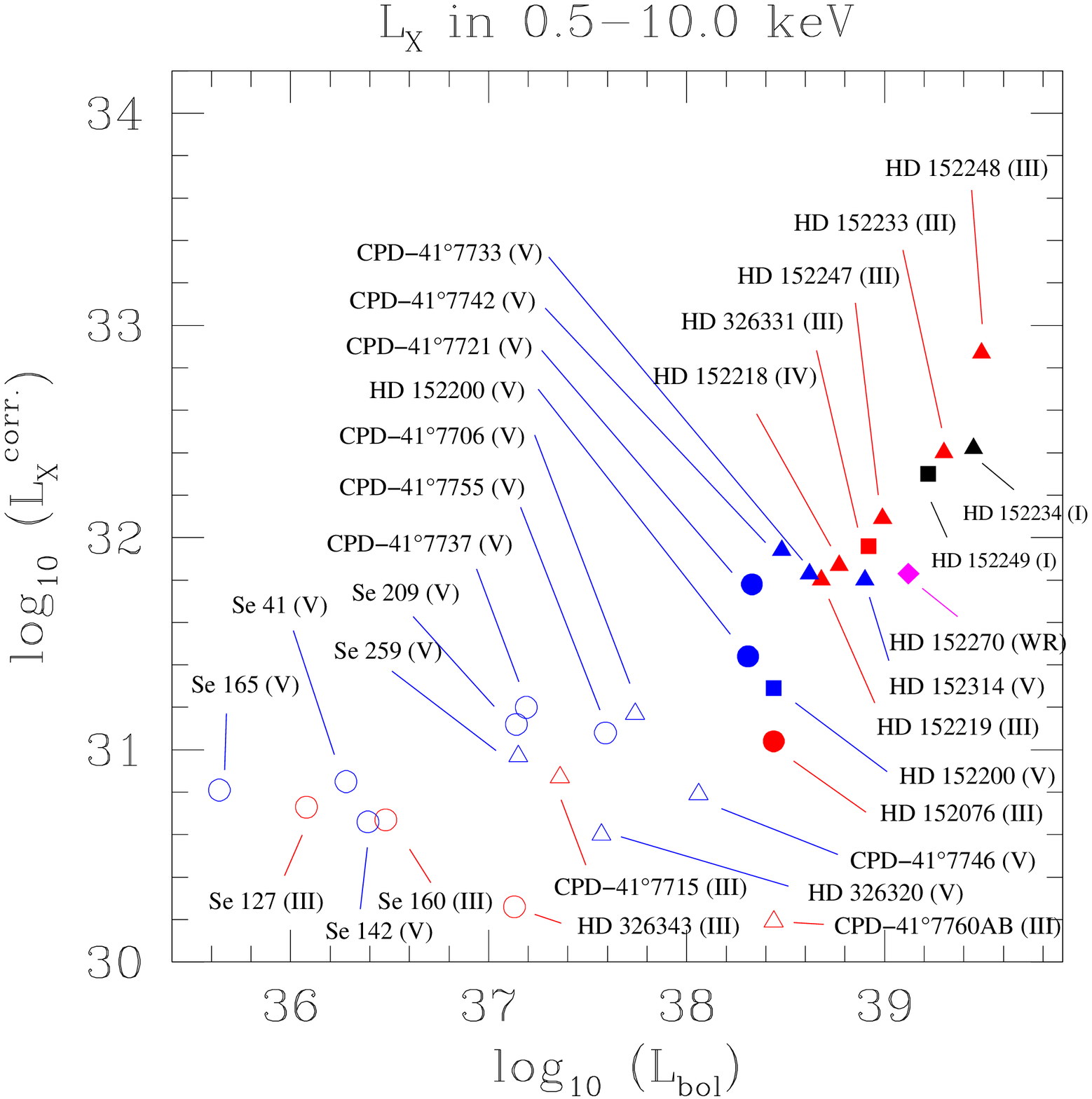}\vspace*{3mm} 
        \includegraphics[width=\columnwidth]{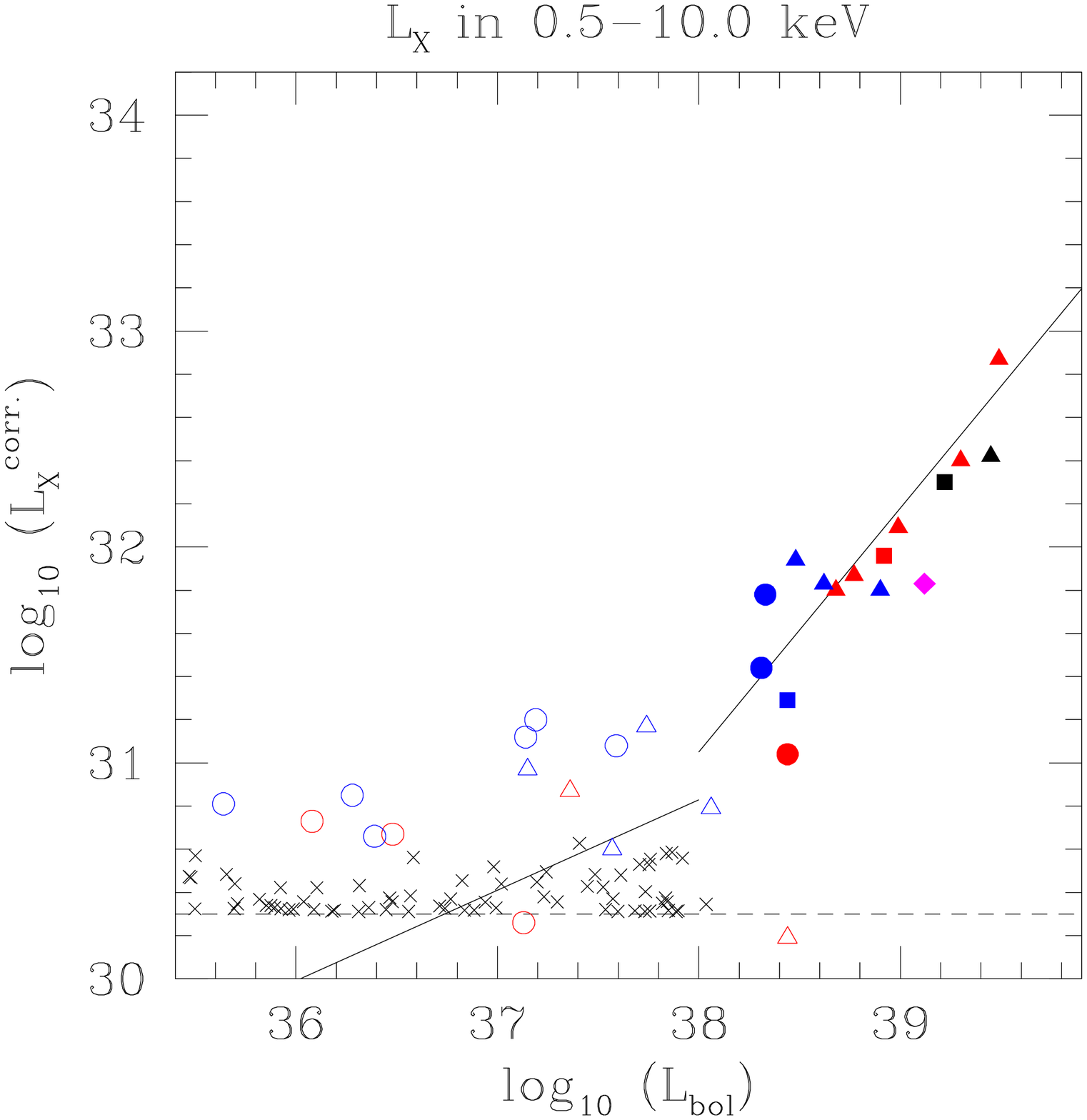}
     \caption{{\bf Top panel}: ISM-absorption corrected X-ray luminosities in the 0.5-10.0~keV band plotted vs. bolometric luminosities. The different symbols indicate the different properties of the sources. Spectral type: O (filled symbols), B (open symbols). Luminosity class: supergiant (I or black), giant (III or red), main sequence (V or blue). Multiplicity: binary (triangles), radial velocity variable (squares), presumably single star (circles). {\bf Bottom panel}: same as the upper panel, the plain lines give the \citet{BSD97} relations obtained in the 0.1-2.0~keV band, respectively for objects with $\log L_\mathrm{X}>38$ (\ergs) and for those below. The dashed line indicates our typical detection limit at a distance of about 5\arcmin\ from the axis for a source with a temperature k$T$ around 0.7-1.5~keV and an ISM column of absorbing hydrogen of $N^\mathrm{ISM}_\mathrm{H}=0.26\times10^{22}$~cm$^{-2}$ (see \citetalias{SGR06}). The crosses indicate the upper limits for the undetected B-type stars (see text).
 }
\label{fig: lxlb}
\end{figure}

   \begin{figure}
     \centering
        \includegraphics[width=\columnwidth]{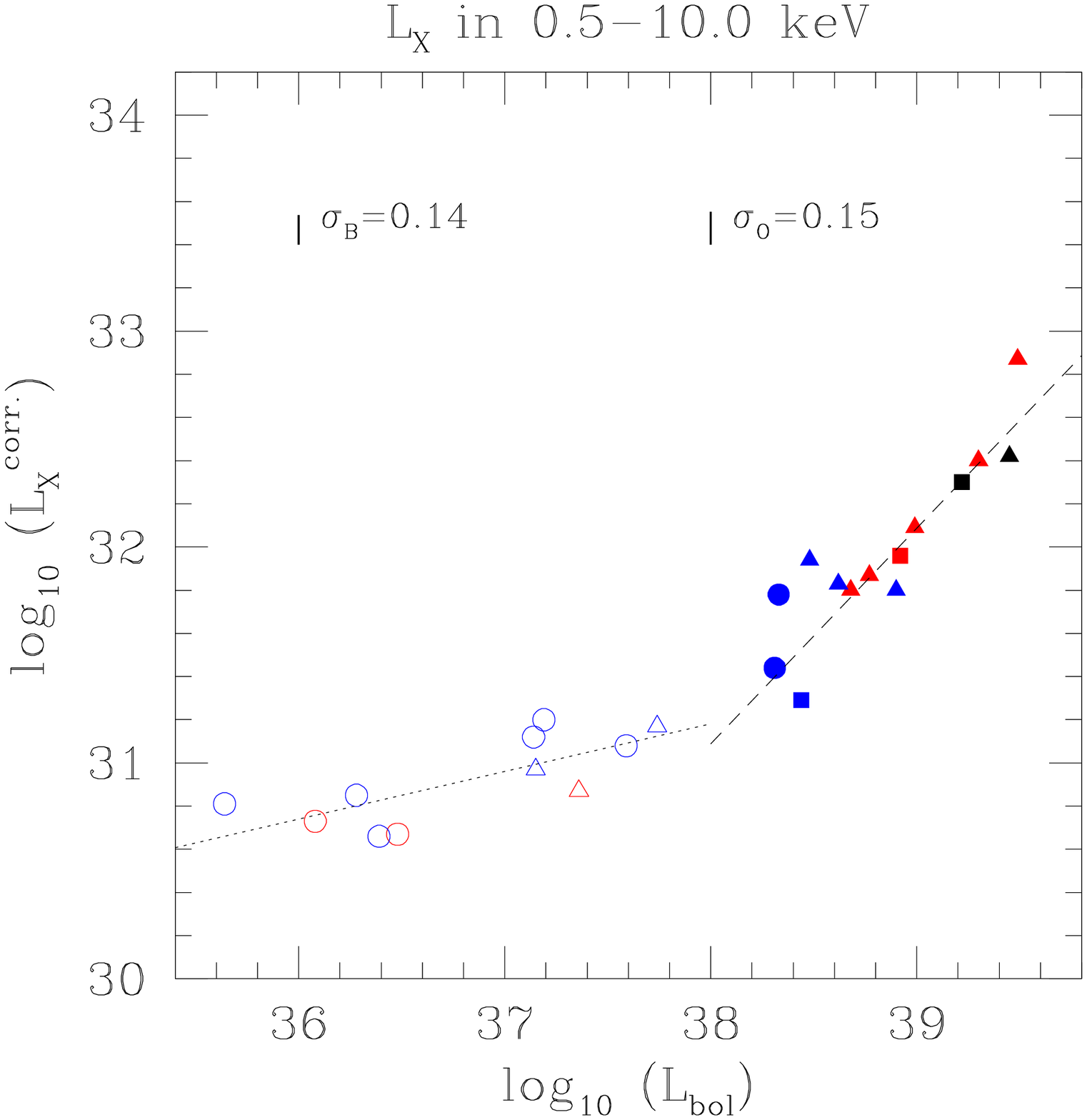}
     \caption{ISM-absorption corrected X-ray luminosities in the 0.5-10.0~keV energy band plotted vs.\ bolometric luminosities. The symbols have the same meanings as in Fig.~\ref{fig: lxlb}. Best-fit linear relations in the \lg\ plane for O (Eq.~\ref{eq: OmW}) and B (Eq.~\ref{eq: B}) stars are indicated by the dashed and the dotted lines respectively. The two vertical bars in the upper part of the graph give the expected 1-$\sigma$ deviation for B ($\sigma_\mathrm{B}$) and O ($\sigma_\mathrm{O}$) type stars.}
     \label{fig: clean}
   \end{figure}

 The absorbing column associated with the lower temperature component is generally very low. The best-fit value might drop to zero so that only an upper limit can be provided. It is known, from past experience, that suppressing the local absorbing column of the soft component can lead to a better fit and to more stable results \citep[see e.g.][]{SSG04, SAR05}. This suggests that the only effective absorption for the related component is due to the ISM. When appropriate this is indicated by a `n.' in Tables \ref{tab: Xspec_1T}, \ref{tab: Xspec_2T} and \ref{tab: Xspec_3T}. The temperature of the second component is usually about 0.7~keV and the associated column of absorbing matter remains moderate, with a mean value about $0.5\times10^{22}$~cm$^{-2}$. We note that the described pattern for the X-ray emission from O-type stars as revealed by the 2-T models is also seen in the results of the 3-T models (Table \ref{tab: Xspec_3T}). The first two components are almost unchanged. An additional high temperature component (with k$T>2$~keV) however describes the higher energy part of the spectrum. This third component has generally an emission measure at least five times lower than the less energetic components  and is probably only seen because the corresponding sources are bright X-ray emitters, which allows us to probe the higher energy tail of their spectrum. Alternatively, the high energy components could come from an unresolved PMS companion located along the same line of sight (though, as stated above, this is rather unlikely). These objects have indeed X-ray spectra typically harder than O-type stars \citep{FeM99}, but should not bias the measured X-ray flux of the O stars as the latter is at least one order of magnitude brighter.

The X-ray spectra of the B-type stars show  quite different properties (see Table \ref{tab: Xspec_2T} and Figs.~2 to 5). Compared to the O stars, the low energy component has a slightly lower temperature but a much larger absorbing column. The main difference however is seen for the high temperature component which is systematically larger than 1~keV. Beyond this spectral scheme, it is obvious that the B-type X-ray emitters are fainter than the O-type emitters and, probably for this reason, we never needed a 3-T model to satisfyingly reproduce their spectrum.


\subsection{\lxlbol\ relationship \label{ssect: lxlbol}}

To derive the intrinsic X-ray luminosities, we corrected the observed fluxes quoted in Tables \ref{tab: Xspec_1T} to \ref{tab: Xspec_3T} for the interstellar reddening. We then computed the corresponding unabsorbed fluxes and, adopting $DM=11.07\pm0.04$, the X-ray luminosities. 
Fig.~\ref{fig: lxlb} presents the $\log L_\mathrm{X}$ vs. $\log L_\mathrm{bol}$ diagram. The different properties of the X-ray emitters (spectral type, luminosity class, multiplicity) and their location in the $\log-\log$ plane are given in the upper panel of the figure. The lower panel gives, in addition, the upper detection limit for the undetected B-type stars, computed according to their respective positions on the detector. These limits were estimated using Eq.~5 and Fig.~6 from \citetalias{SGR06} and refer to a source with a temperature about 0.7-1.5~keV and a typical ISM reddening corresponding to an equivalent column of neutral hydrogen of $0.26\times 10^{22}$~cm$^{-2}$. We emphasize that no additional (local) absorption column is considered in this typical model. \\

Clearly the O- and B-type stars show different behaviours. The former ones are brighter (as expected) and present a steeper \lxlbol\ relation than the latter. The B-type stars and the late O-type stars also seem to present a larger dispersion around the observed trends. However, the impression of dispersion towards lower $L_\mathrm{X}$ mainly comes from five points, namely HD~152076 for the O-type stars and \cpd7760, \cpd7746, HD~326320 and HD~326343 for the B-type sample. These sources correspond to five out of the seven OB sources with an off-axis distance over 10\arcmin. The other two off-axis sources are  HD~152235, a B1Ia star whose X-ray emission could not be disentangled from a brighter neighbour (and which is thus not plotted in the current diagram; see also Fig.~\ref{fig: hd235} in Sect.~\ref{ssect: Bindiv}), and \hd247, an X-ray bright O9~III star. With the exception of the latter object, all the studied sources with an off-axis angle larger than 10\arcmin\ display a smaller X-ray luminosity compared to the general trend in the sample. Further inspection of the results of Tables \ref{tab: Xspec_1T} to \ref{tab: Xspec_3T} reveals that the X-ray fluxes from these five sources were all obtained with single temperature models. These were characterized by  k$T$ about 0.25~keV for the O star and about 0.5-0.6~keV for the four other B stars. Interestingly, the obtained k$T$ corresponds to the lower temperature component of the 2-T models. We caution that, because these sources are located in the outer parts of the detector, where the sensitivity significantly decreases, only the softest part of the spectrum produces a significant number of photons. The harder part of the spectrum provides too faint a contribution for a 2-T model being used, which probably yields an underestimation of the observed flux in the 0.5-10.0~keV band. HD~152247 being about one order of magnitude  brighter in the X-rays, the higher energy part of its spectrum remains relatively well constrained. The obtained spectrum is indeed of good quality and a 2-T model is definitively required to properly reproduce it.  
To preserve as much as possible the homogeneity of our sample, both in terms of location in the cluster and in terms of the hypotheses that underlie the analysis, we have restrained our discussion  to the results obtained with the 2-T and 3-T models.  \\


   \begin{figure*}
     \centering
        \includegraphics[width=.32\textwidth]{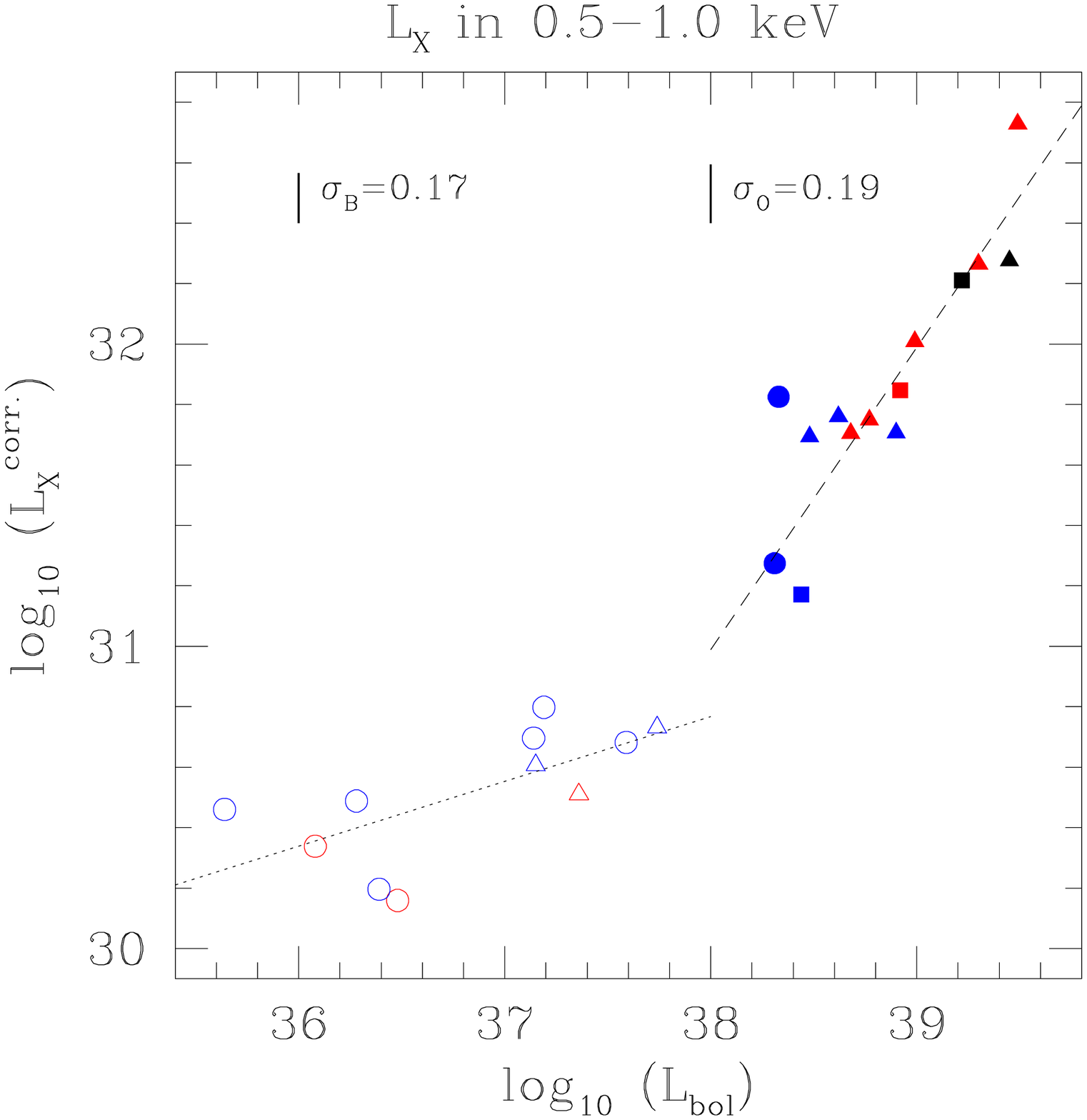}
        \includegraphics[width=.32\textwidth]{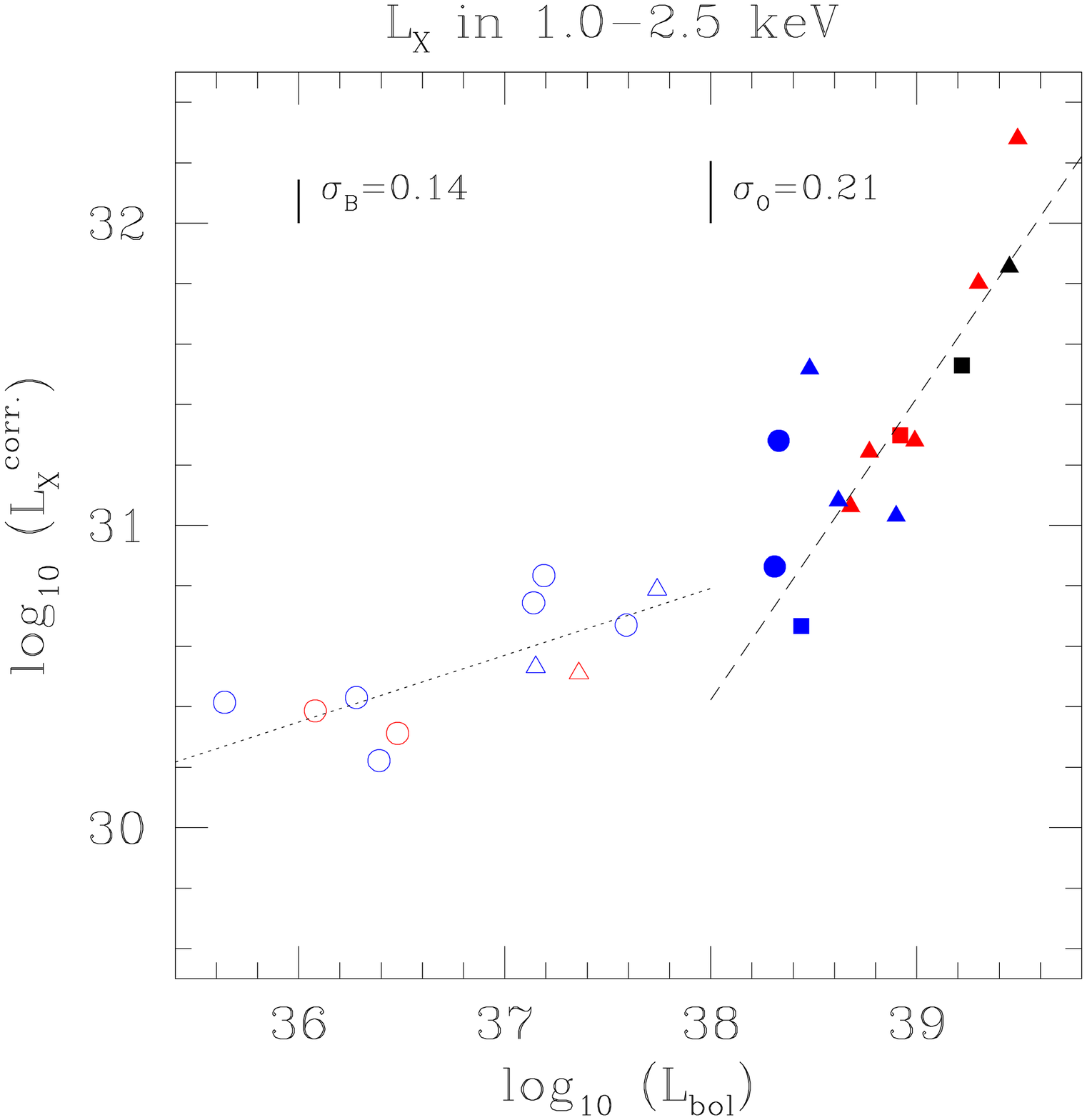}
        \includegraphics[width=.32\textwidth]{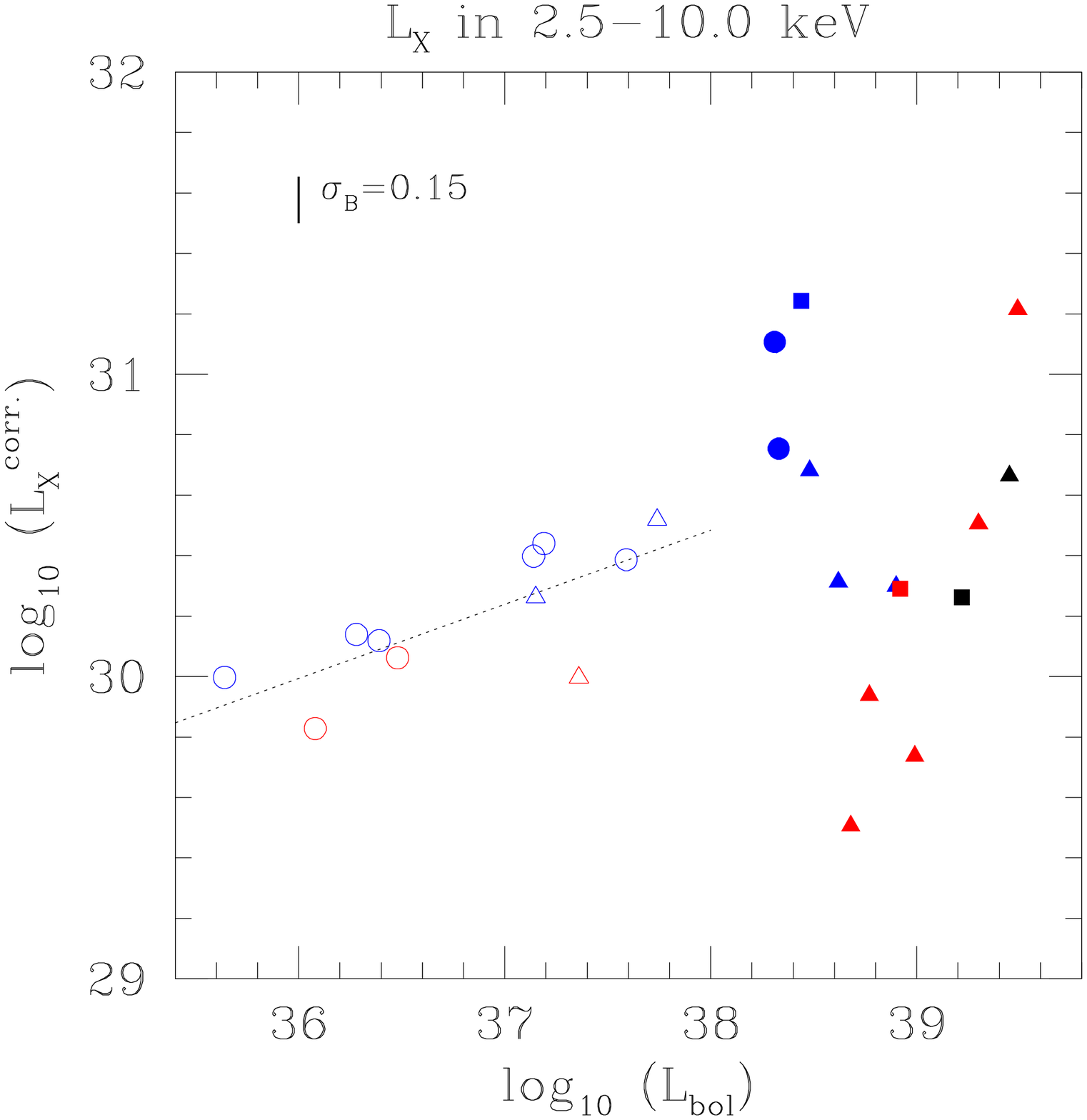}
     \caption{ISM-absorption corrected X-ray luminosities in different energy bands plotted vs. bolometric luminosities. The symbols have the same meanings as in Fig.~\ref{fig: lxlb}. Best-fit linear relations in the \lg\ plane are indicated by dotted and dashed lines respectively for B- and O-type stars.}
     \label{fig: clean_eb}
   \end{figure*}

Before deriving the \lxlbol\ relationships for our sample, we would like to draw the attention of the reader to the following point. In the rest of this paper, we will make a difference between a scaling law (SL):
 $$L_\mathrm{X}=K\ L_\mathrm{bol},$$
 where $K$ is the scaling constant, and a power law (PL): 
 $$L_\mathrm{X}=K\ L_\mathrm{bol}^\Gamma,$$
 where $\Gamma$ is the power law index. In the \lg\ plane, both relations yield a straight line: 
$$\log L_\mathrm{X} = \log L_\mathrm{bol} + \log K$$
 for the scaling law, and 
$$\log L_\mathrm{X} = \Gamma\ \log L_\mathrm{bol} + \log K$$
 for the power law. We emphasize that the former expression has a slope equal to unity in the \lg\ plane and is equivalent to the statement $\log \left(L_\mathrm{X}/L_\mathrm{bol}\right)=cst$, with $cst=\log K$. The earliest works on the canonical relation mostly used scaling relations while, later on, power laws were used to provide an additional degree of freedom to the fitted model. In the following, both kinds of models will be tested and their results will be compared.

\subsubsection{X-ray emission from O-type stars ($L_\mathrm{bol}>10^{38}$~\ergs)}

As seen in Figs. \ref{fig: lxlb} and \ref{fig: clean}, the O-type stars mostly concentrate on a line in the \lxlbol\ plane. As a first step in the analysis, we computed the linear-correlation coefficient $r$ (Table \ref{tab: plo}). We then estimated  the probability  for an uncorrelated parent distribution to yield a value of $r$ equal or larger than the observed value \citep[see e.g.][]{Bev69}. For all the considered O star samples (see below), we could reject this hypothesis at a significance level of  0.001.
Assuming first that $L_\mathrm{X}/L_\mathrm{bol}=cst$,  a least-square fit performed on the 14 objects (sample A) yields: 

\begin{equation}
\log \left(L_\mathrm{X}/L_\mathrm{bol}\right)_\mathrm{A}= -6.865 \pm 0.186 \label{eq: Oall} 
\end{equation}
with a typical dispersion of the X-ray luminosity of about 55\% around Eq.~\ref{eq: Oall}. However, we note that most of the dispersion comes from four points. Among them, \hda\ and \cpd7742\ are known to display a significant extra-emission associated with a wind interaction phenomenon occurring in both systems \citep{SSG04, SAR05}. This extra-emission, averaged over the orbital cycle may account for at least one third of the observed flux (see individual notes for more details).
Rejecting these two systems from the fit (sample B), we now obtain: 

\begin{equation}
\log \left(L_\mathrm{X}/L_\mathrm{bol}\right)_\mathrm{B}= -6.912 \pm 0.153 \label{eq: OmW}
\end{equation} 
with a typical 1-$\sigma$ deviation corresponding to about 40\% on \lx\ (see Fig.~\ref{fig: clean}).
 Beyond the two colliding wind systems, the remaining dispersion is mainly due to \hd200 and HD~326329. It is interesting to note that, with a respective spectral type of O9.5 and O9.7,  these two presumably single dwarfs are the two coolest O-type stars of our sample. Hence, further excluding the stars with a sub-spectral type later than O9 (sample C), the best fit values are almost unchanged:

\begin{equation}
\log \left(L_\mathrm{X}/L_\mathrm{bol}\right)_\mathrm{C}= -6.925 \pm 0.087 \hspace*{2mm}.\label{eq: OmV}
\end{equation}

The residuals drop this time to 20\%. Henceforth, with the exception of the very late O-type main-sequence stars and of the two identified systems with strong wind interaction effects, the  X-ray luminosity  of the other objects displays a very limited dispersion around the canonical relation. Among these ten stars, dwarfs, giants and supergiants are found, as well as presumably single stars and binaries\footnote{The X-ray contribution of a wind interaction to the total emission of the latter binaries is expected to be very limited (see Sect.~\ref{sect: indiv}). Their X-ray emission should thus reflect very closely the intrinsic emission of their components and, indeed, their X-ray luminosities do not deviate from the canonical relation.}. In Sect.~\ref{sect: indiv}, we pay a more particular attention to each of the X-ray emitters. \\
\begin{table*}
\caption{Comparison between best-fit \lxlbol\ scaling law (SL) and power law (PL) models, computed in the 0.5-10.0~keV energy bands (upper part of the table) and in the soft (S$_\mathrm{X}$) and intermediate (M$_\mathrm{X}$) energy ranges (bottom part). The first two columns specify the O star sample used (see text) and the number $N$ of stars in the sample. The third column provides the $\log L_\mathrm{X} - \log L_\mathrm{bol}$ correlation coefficient $r$. The next two columns indicate the best-fit scaling laws, the reduced $\chi^2$ and the number of degrees of freedom (dof) of the fit. Cols. 6 and 7 yield the same information for the best-fit power laws. Note that, in Cols. 5 and 7, the $\chi^2$ statistics have not been normalized by the intrinsic variance of the data. The last column gives the values of the statistics $F_\chi$.}
\label{tab: plo}
\begin{tabular}{c c c c c c c c}
\hline
Sample & $N$ & $r$ & SL (Eq.~\#) & \chin\ (dof) & PL & \chin\ (dof)  & $F_\chi$ \vspace*{1mm}\\
\hline
A & 14 & 0.89 & Eq.~\ref{eq: Oall} & $3.46\times 10^{-2}$ (13)               & \hspace*{2.4mm}$\log L_\mathrm{X} = 0.91(\pm0.13) \log L_\mathrm{bol} - 3.6 (\pm 5.1)   $ & $3.62\times 10^{-2}$ (12)               & 0.4 \\
B & 12 & 0.91 & Eq.~\ref{eq: OmW}  & $2.34\times 10^{-2}$ (11)               & \hspace*{2.4mm}$\log L_\mathrm{X} = 0.85(\pm0.12) \log L_\mathrm{bol} - 1.2 (\pm 4.7)   $ & $2.24\times 10^{-2}$ (10)               & 1.5 \\
C & 10 & 0.97 & Eq.~\ref{eq: OmV}  & $7.56\times 10^{-3}$ \hspace*{1.5mm}(9) & \hspace*{2.4mm}$\log L_\mathrm{X} = 0.88(\pm0.08) \log L_\mathrm{bol} - 2.2 (\pm 3.1)   $ & $6.58\times 10^{-3}$ \hspace*{1.5mm}(8) & 2.3 \\
\hline
B & 12 & 0.86 & Eq.~\ref{eq: Os}   & $3.75\times 10^{-2}$ (11)               & \hspace*{1.mm}$\log L_\mathrm{X, S} = 0.80(\pm0.15) \log L_\mathrm{bol} + 0.6 (\pm 5.9)$ & $3.53\times 10^{-2}$ (10)              & 1.7 \\
B & 12 & 0.85 & Eq.~\ref{eq: Om}   & $4.22\times 10^{-2}$ (11)               & $\log L_\mathrm{X, M} = 0.75(\pm0.16) \log L_\mathrm{bol} + 0.6 (\pm 6.2)$ & $3.95\times 10^{-2}$ (10)              & 1.8 \\
\hline
\end{tabular}
\end{table*}

As a next step, we computed the X-ray luminosities of the sample stars in different energy bands. Plotted against the bolometric luminosities (Fig.~\ref{fig: clean_eb}), we again observed linear relations in the \lg\ plane, both in the soft (S$_\mathrm{X}$: 0.5-1.0~keV) and medium (M$_\mathrm{X}$: 1.0-2.5~keV) bands, while a large dispersion is seen at higher energies (H$_\mathrm{X}$: 2.5-10.0~keV). In the S$_\mathrm{X}$ and M$_\mathrm{X}$ bands, the relative position of the stars remains very similar. The two colliding wind systems display a larger deviation in the intermediate band than in the soft band. This is to be expected as the X-ray emission produced in the wind-interaction region is typically harder than the intrinsic emission from O-type stars, which drops significantly beyond 1-1.5~keV. Using the sample B, best-fit relations in the soft (0.5-1.0~keV) and intermediate (1.0-2.5~keV) energy ranges yield respectively:

\begin{equation}
\log \left(L_\mathrm{X,S}/L_\mathrm{bol}\right)_\mathrm{B}= -7.011 \pm 0.193 \label{eq: Os}
\end{equation}
and 
\begin{equation}
\log \left(L_\mathrm{X,M}/L_\mathrm{bol}\right)_\mathrm{B}= -7.578 \pm 0.205 \label{eq: Om}
\end{equation}
with a typical dispersion on \lx\ of about 55\% and 60\% respectively.  The hard band shows a much larger dispersion and further suggests a slightly different behaviour for main-sequence stars compared to more evolved objects. The former ones are about at least one order of magnitude brighter than the giants of a similar spectral type. There also seems to be a steep trend in  the giant and supergiant population and their fluxes in the hard band indeed increase by more than one order of magnitude towards larger bolometric luminosities. We caution however that the measured X-ray flux above 2.5~keV is probably little constrained, except for the brightest X-ray emitters, and that the present results for the hard band probably require confirmation.

As mentioned earlier, we also adjusted power laws to the data by means of a linear regression performed in the \lg\ plane. The best-fit relations are quoted in Table \ref{tab: plo}. These only yield a marginal improvement of the residuals and the dispersion of the data around the power law relations is virtually unchanged. To test the judiciousness of including this additional degree of freedom in the model, we  used a $F_\chi$-test as described in e.g.\ \citet{Bev69}.  The $F_\chi$ statistic is given in Table \ref{tab: plo} and was obtained from the equation:
\begin{equation}
F_\chi=\frac{\chi^2_\mathrm{SL}-\chi^2_\mathrm{PL}}{\chi^2_\mathrm{PL}/(N-2)}
\end{equation}
where $N$ is the number of points in the considered sample. Clearly, the $F_\chi$-test does not indicate any significant improvement of quality of the fit, even when adopting a significance level as large as 0.1. In consequence, we consider that the power law model does not provide any significant improvement compared to the simpler scaling law. In the following, we have adopted the canonical \lxlbol\ relation given by Eq.~\ref{eq: OmW}.\\

\subsubsection{X-ray emission from B-type stars ($L_\mathrm{bol}<10^{38}$~\ergs)}

The present sample of B-type stars is strongly dominated by non-detections for which only an upper limit on the actual X-ray luminosity can be obtained. To account for this, we have attempted to apply various survival analysis techniques \citep[e.g.][]{IFN86}. Unfortunately, we could not achieve consistent results, probably because of the much larger number of upper limits compared to the actual measurements. Therefore we have chosen to restrain our analysis to the detected objects only, making thus the implicit hypothesis that the latter stars form a specific population, distinct from the undetected B stars.

As mentioned in Sect.~\ref{ssect: Xspec}, the detected B-type stars have a quite harder emission (k$T_2>1.0$~keV) compared to  the O-type stars. They are however much more luminous in the 0.5-10.0~keV band than predicted by the \citet{BSD97} relation (Fig.~\ref{fig: lxlb}). This could result from the fact that the higher energy part of the B-type spectrum falls outside the observable energy range of the \rosat\ satellite.
 Like the O-type stars, their X-ray and bolometric logarithmic luminosities are apparently still linked by a linear relation. The  linear-correlation coefficients $r$ are about 0.75, which rejects the null hypothesis of such a high value occurring by chance from an uncorrelated distribution at a significance level of 0.01. Only the $r$ value computed for the S$_\mathrm{X}$ band is slightly lower ($r=0.68$) and yields the rejection of the null hypothesis only if we adopt a significance level of 0.05.
It is obvious, from Figs. \ref{fig: lxlb} to \ref{fig: clean_eb}, that the slope of a putative relation is significantly different from unity and indeed the $F_\chi$ test described above indicates that the power law provides a very significant improvement of the quality of the fit compared to a simple scaling law. From our selected sample of 11 B-type stars, we obtain:
\begin{equation}
\log L_\mathrm{X} = 0.22 (\pm0.06) \log L_\mathrm{bol} + 22.8 (\pm 2.4)  \label{eq: B}
\end{equation}
with a typical dispersion around Eq.~\ref{eq: B} of  0.14 in the \lg\ plane, thus corresponding to about 37\% in the X-ray luminosities.

From Fig.~\ref{fig: clean_eb}, the emission of the detected B-type stars in the \lg\ plane also follows a linear relation in  the different sub-energy bands considered. By opposition to the behaviour of O-type stars, this linear behaviour holds in the harder band too. However, we note that the softest part ($<5$~keV) of the hard band is responsible for most of the flux. The latter remains indeed very limited above this value (see e.g.\ Fig.~4).  From linear regressions, we obtain:

\begin{equation}
\log L_\mathrm{X, S} = 0.21 (\pm0.08) \log L_\mathrm{bol} + 22.6 (\pm 2.8)  \label{eq: Bs}
\end{equation}

\begin{equation}
\log L_\mathrm{X, M} = 0.22 (\pm0.07) \log L_\mathrm{bol} + 22.4 (\pm 2.5)  \label{eq: Bm}
\end{equation}

\begin{equation}
\log L_\mathrm{X, H} = 0.24 (\pm0.07) \log L_\mathrm{bol} + 21.1 (\pm 2.6)  \label{eq: Bh}
\end{equation}
with typical dispersions of respectively  0.17 (47\%), 0.14 (39\%) and 0.15 (42\%) around these relations.
 Before presenting a more detailed discussion of the derived \lxlbol\ relations and comparing these with earlier works (see  Sect.~\ref{sect: discuss}), the next section will first review the individual properties of the early-type X-ray emitters.

\section{Individual objects} \label{sect: indiv}

This section aims at providing specific information on the various early-type X-ray emitters detected in the \xmm\ FOV. In particular,  we haverm -R S investigated their variability on the different time-scales allowed by our data set. Using the \eml\ net count rates obtained in Sect.~\ref{sect: obs}, we performed a $\chi^2$ test of hypothesis with a significance level of 0.01 and we actually tested the null hypothesis of a constant count rate throughout the six pointings of the \xmm\ campaign. Nine sources displayed consistent indications for variability in all the \epic\ instruments. These are noted `var.'  after their name in the individual sections here below. In addition, \cpd7715 and \cpd7737 gave consistent results in all the \epic\ instruments with a significance level of 0.1 only. They are noted `var?'. Finally, HD~326329 presented significant variations at the 0.01 level but was probably contaminated by a neighbouring strong source (X\#324). The latter displayed a strong flare that is also seen in the HD~326329 count rates. It is thus difficult to determine whether, in addition to the contaminating component, the HD~326329 source displays intrinsic variation of its flux or not. 

To search for intra-pointing variability (i.e.\ on time scales shorter than the various exposure durations), we further applied the Kolmogorov-Smirnov and probability of variability tests described in \citet{SSG04} for the case of \hd248. While the latter non-parametric tests do not make any a priori hypothesis on the form of the variability, they do not allow to correct for the background. We thus also built background and GTI corrected light curves using various bin sizes ranging from a few tens of seconds to a few kiloseconds and we applied similar $\chi^2$ tests as previously done for \hd248. As in the case of inter-pointing variability, we adopted a significance level of 0.01 for the various tests. For the O-type stars, intra-pointing variability is only seen for the two strong colliding wind binaries while, for B-type stars, the tests reveal several flaring sources. Thanks to the amplitude of the flares, the latter sources had however already been detected as variable by the inter-pointing variability study.

\subsection{O-type stars} \label{ssect: Oindiv}
\begin{enumerate}

\item[-] {\bf \hd076} is located at 13\arcmin\ N-NW of  the cluster core. It lies outside the \pn\ detector FOV and is thus only seen in the two \mos\  instruments. It presents a constant flux throughout the campaign.  The quality of the obtained spectra is definitively too low to constrain a 2-T model. We suspect the adopted single temperature model to bias the computed flux in the 0.5-10.0~keV band. This source was thus not included in the derivation of the \lxlbol\ relationship. 
\\
\item[-] {\bf \hd200} is the latest O-type star of our sample. It is reasonably bright and well isolated in the X-ray images and its X-ray flux remains constant. We considered it to be reasonably  well constrained. \hd200 is however one of the O-type  stars that deviate the most from the derived \lxlbol\ relation. We have not been able to identify a cause that could possibly bias the determination of its flux and we consider the observed deviation to be significant.
\\
\item[-] {\bf \hd218} displays a slight modulation of its flux by about 20\% but its X-ray spectral properties remain almost unchanged \citep{phd}. The observed modulations could be linked to a possible wind-wind collision occurring in this O+O system, but the latter authors note that second order effects, such as radiative inhibition, might actually govern the wind interaction properties in this system. However, the \hd218 X-ray flux remains probably dominated by the intrinsic emission from the stars.
\\
\item [-] {\bf \hd219} (var.) presents a slight modulation of its flux \citep{SGR06_219} which could be related to a similar wind interaction as in \cpd7742. Except for the longer orbital period of \hd219, the two systems are indeed very similar. The amplitude of the variation seems however  much more limited than in \cpd7742. This probably results from the longer period of the system, yielding thus a larger separation between the components and a subsequent dilution of the wind material that enters the interaction. \hd219 is the third O-type eclipsing binary known in the cluster \citep{SGR06_219}
\\
\item[-]{\bf \hd233} is most probably a long period binary \citep[$P \sim $ months,][]{phd}, but its orbital properties are still not constrained. The \hd233 X-ray flux presents an apparently constant level throughout the campaign.
\\
\begin{figure}
\centering
\includegraphics[width=\columnwidth]{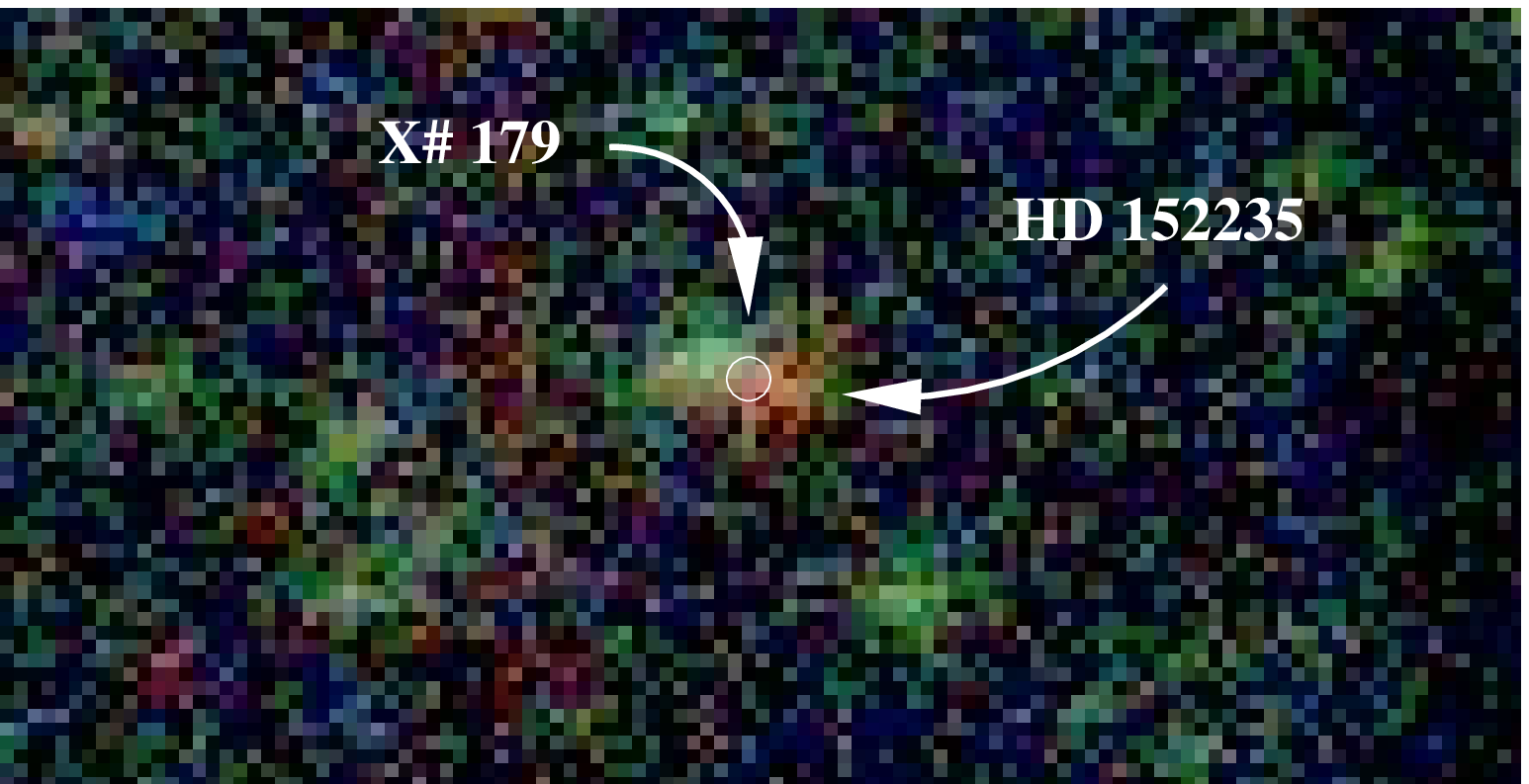}
\caption{{\bf \hd235:} \epic\ three-colour image: red, S$_\mathrm{X}$ band; green, M$_\mathrm{X}$ band; blue, H$_\mathrm{X}$ band. The location of X\#179, indicated by the 3\arcsec\ radius circle, is clearly biased by the presence of two close, unresolved sources that can only be distinguished in this image thanks to their different spectral properties.  The supergiant B star \hd235 is clearly associated with the soft emission. The image appears in colour in the electronic edition of the paper.}
\label{fig: hd235}
\end{figure}
\item[-]{\bf \hd234} (var.) is reminiscent of \hd233. The system  is most probably a long period ($P \sim $ months) binary \citep{phd}, but its orbital properties are poorly constrained so far. Its X-ray flux slightly decreased  throughout the \xmm\ campaign with an average slope of about $-4.3\times 10^{-3}$~\cnts\ per day. This corresponds to a reduction from about $75\times 10^{-3}$ to about $55\times 10^{-3}$~\cnts\ between the first and the sixth observations. This could be due to a wind interaction, between the two components, whose strength is varying because e.g.\ of the changing separation between the stars. However, a more precise analysis  requires first to constrain the orbital properties of the system. As for HD~152233, an ongoing monitoring campaign should soon provide this missing piece of the puzzle.
\\
\item[-]{\bf \hd235} (not resolved) is a B1Ia star in the southern part of the field. The cross-correlation did not associate this object with an X-ray source. The three-colour image (Fig.~\ref{fig: hd235}) however clearly reveals that \hd235 is a very soft X-ray emitter. The source detection has actually been biased because of a close, harder emitter. Due to this close neighbour, the \hd235 spectrum can unfortunately not be reliably extracted and the source was thus not included in the present analysis.\\

\item[-]{\bf \hd247} is located in the outer region of the field. It is most probably an SB1 binary \citep{phd}, but its period is largely undetermined. It could however be of the order of one year. The X-ray flux seems to remain constant from one pointing to the other as well as within each pointing.\\

\item[-] {\bf \hda} (var.): the colliding wind binary \hda\ is the brightest X-ray source in the FOV \citep{SSG04} and presents phase-locked modulations of its X-ray flux due to the varying strength of the shock in this eccentric system. The X-ray data have been thoroughly analysed in \citet{SSG04}. To provide averaged properties for this peculiar system, we adjusted a 3-T model to its spectrum. Unfortunately, we could not get \chin\ values better than 2.4. This might reflect the changing spectral properties of the system and the impossibility for \mek\ models to  reproduce, in this case, an averaged emission level. Because of the \chin\ $ > 2$,  the \xspec\ software did not allow to explore the parameter space to search for a better local minimum. For the same reason, we could not directly compute the limits of the 90\% confidence intervals. In Table \ref{tab: Xspec_3T}, the quoted intervals were estimated from the 1-$\sigma$ error bars that were converted into 90\% confidence intervals assuming a Gaussian distribution of the uncertainties. We note however that the obtained flux in the 0.5-10.0·~ keV band is very similar to the average of the fluxes observed during the six individual pointings \citep{SSG04}. This indicates that the present result is quite consistent with the previous analysis. Using the newly derived \lxlbol\ relation of Eq.~\ref{eq: OmW} and the individual luminosities, we expect $\log L_\mathrm{X}=32.58$. We observed, on average, $\log L_\mathrm{X}=32.87$, which is about twice larger. 

In \citet{SSG04}, we have modelled the X-ray emission expected from the wind-wind interaction using hydrodynamical simulations. Once diluted by the intrinsic emission from the two stars, the modelling qualitatively reproduced the observed modulations. However, the predicted X-ray  flux was about a factor two too faint compared to the observed emission level of the system. At the time, we attributed this difference to the observed dispersion around the \citet{BSD97}'s canonical relation. From the present results, this dispersion is presumably much more limited than previously expected. Therefore, the \hda\ system probably deserves further theoretical work to quantitatively explain the observed modulations and the exact shape of its light curve. \\


\item[-] {\bf \hd249} presents a strong increase of its flux during Obs.~2. This increase is seen in the three energy  bands, but is most prominent in the intermediate  and in the hard  bands. The corresponding images reveal a nearby flaring source located at about 6\arcsec\ S-W of \hd249 (thus well within the extraction region). While a slight contamination is still detected in the hard band during Obs.~3, the count rates measured during Obs~1, 4, 5 and 6 remain in good agreement and no additional intra- or inter-pointing variability could be found. Both the \hd249\ spectral fit given in Table \ref{tab: Xspec_2T} and the derived fluxes and luminosities rely only on the four pointings with no contamination.
\\

\item[-] {\bf \hd314} is a relatively isolated bright X-ray emitter whose X-ray flux seems to remain constant.\\

\item[-]{\bf HD~326329} (contaminated?) is located in a relatively crowded region of the cluster. The count rates obtained via the SAS task {\it emldetect}  are clearly affected by the bright X-ray source X\#324 that displayed a flare during Obs.~4. We re-derived the count rates in a more limited extraction radius of 7\farcs6 but obtained mitigated results. Clearly the derived fluxes for this object should be regarded with caution. It is interesting to note that, after the two wind interaction systems, HD~326329 presents the largest deviation from Eq.~\ref{eq: OmW}. From the above remark, this dispersion is probably an artefact and might not be linked to any physical process.  \\

\item[-]{\bf HD~326331} reveals particularly broad lines displaying slight profile variations in its optical spectrum. Some authors \citep{HCB74, LM83, GM01} have suggested that it could actually be a binary system. From our work \citep{phd}, we believe that the binarity is not confirmed and that the profile variations are probably not, or not only,  due to the presence of a companion. The X-ray flux of HD~326331 remained constant over the duration of the campaign.\\

\item[-] {\bf \cpd7721} is a visual double star with components, labelled `p' and `s', separated by $\sim5.8\arcsec$. Other denominations for these components are SBL~350 and 351 or BVF~12 and 27.  The X-ray source X\#251 is clearly associated with the O-type `p' component. Abnormally large residuals are seen at the position of another star, SBL~358 ($V=15.89$), a neighbouring object, but actually there seems to be no detected X-ray emission associated with the B-type star \cpd7721s. The {\it emldetect} count rates of the 'p' component show some deviating points, especially at Obs.~2 for the \mos2 instrument and at Obs.~4 for the \pn. These deviating points are however not confirmed by the other instruments. We re-derived the count rates using the same extraction region as the one adopted for the spectra and we found that the X-ray flux from \cpd7721p is probably constant throughout the campaign. We note that the extracted spectra might be contaminated by the residual X-ray emission located at $\sim$8\arcsec\ from \cpd7721p and probably associated with SBL~358.  The contamination is however probably very low and limited to the soft part of the spectrum, thus in the energy range where \cpd7721p displays the strongest X-ray emission.
\\

\item[-] {\bf \cpd7733} presents an approximatively constant X-ray flux throughout its orbit \citep{phd} and indeed, no significant X-ray overluminosity from a possible wind interaction is expected to be found at our detection threshold. 
 \\

\item[-] {\bf \cpd7742} (var.) is the second known SB2 eclipsing early-type binary in \ngc\ and has a period close to 2.4~d \citep{SHRG03}. It is about 3 times brighter in the X-rays than expected from Eq.~\ref{eq: OmW}. \citet{SAR05} presented the \mos\  X-ray light curve of the system, almost fully covering the orbital cycle with a time bin of 1~ks. We showed that an extra-emission component is associated with the inner face of the secondary. When the orientation of the line of sight is favourable, this component could be responsible for an increase of the X-ray flux by about a factor of 2. These features were interpreted as the primary wind crushing on (or nearly on) the secondary surface. Alternatively, the wind-photosphere interaction could be altered leading to a wind-wind interaction region located close to the secondary surface. Even at its lowest flux level, the logarithmic X-ray luminosity  is still about 0.3 above the one expected for the intrinsic emission of this O+B binary.\\

\end{enumerate}

\subsection{B-type stars} \label{ssect: Bindiv}
Before reviewing the individual B-type emitters, we note that only about 20\% of the B-type stars are associated with an X-ray source in the catalogue of \citetalias{SGR06}. Among these,  four sources are apparently single and, at our detection threshold, do not display significant variations. These are Se~127 ($\equiv$ SBL~746, B8III/IV), Se~142 ($\equiv$ SBL~748, B4V), Se~160 ($\equiv$ SBL~593, B3III), and HD~326343 (B3III). These stars will not be discussed any further.

\begin{figure}
\centering
\includegraphics[width=\columnwidth]{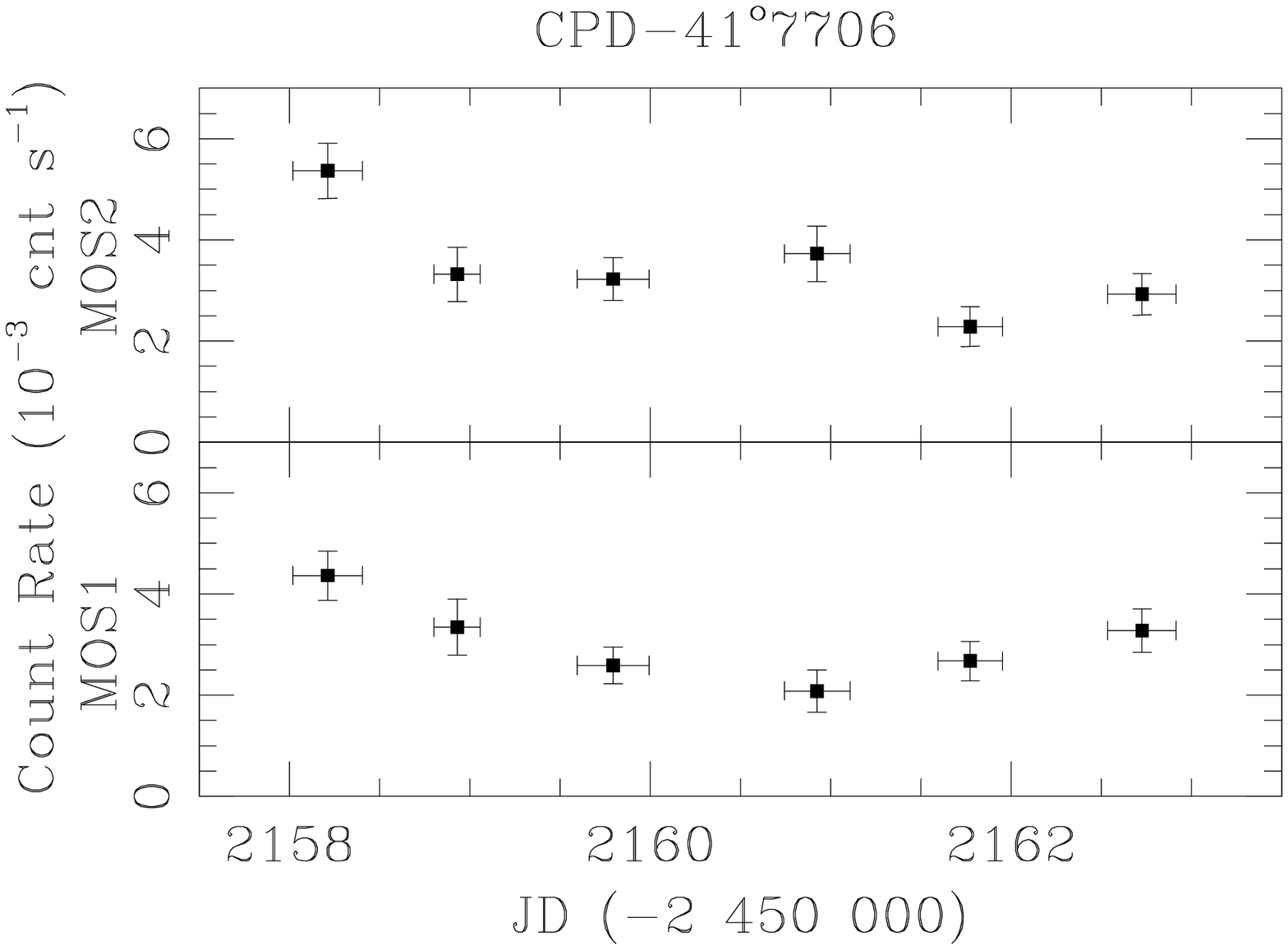}
\caption{{\bf \cpd7706} background corrected \epicmos\ light curves (in the range 0.5--10.0~keV) during the \xmm\ campaign. The average count rates during each pointing are given. The vertical bars are the 1-$\sigma$ deviations while the horizontal bars give the duration of each pointing.}
\label{fig: cpd77706}
\end{figure}

\begin{enumerate}
\item[-]{\bf HD~326320} is an SB2 binary located in the outer regions of the FOV.  At our detection threshold, we do not detect any significant variations but the count rates are quite low. Its spectrum is well described by a single temperature model.
\\
\item[-]{\bf \cpd7706} (var.) is a B1+B1 binary \citep{LM80, GM01}. \citet{phd} reported that the \halph\ and all the \hea\ lines from \l4920 to \l7100 display mixed absorption plus emission. All these lines have the same profile  with a narrow absorption superposed on a broader emission which results in line profiles reminiscent of the Be spectral signature. Because the component signatures are blended, it is not possible to decide whether both components or just one of them is actually a B1Ve star. The source presents a smooth variation of its X-ray flux  (Fig.~\ref{fig: cpd77706}), but no sign of flaring activities could be found for this system.
\\
\item[-]{\bf \cpd7715} (var?) is reported as a spectroscopic binary by \citet{Rab96} and as a \bcep-type star by \citet{ASK01}. There might be an increase of the detected flux during Obs.~2 but a $\chi^2$ test with a  significance level of 0.01 did not allow us to reject the null hypothesis of constant count rate throughout the six pointings. 
\\
\item[-]{\bf \cpd7737} (var?) is a slowly pulsating B2V star \citep{ASK01}. There is a slight increase ($\sim$40\%) of the count rate during Obs.~5, but because of the large error bars, it might not be significant. A $\chi^2$ test with a  significance level of 0.01  did not allow us to reject the null hypothesis of constant count rate throughout the six pointings. Note that another optical source, SSB06~\#3549 ($V=16.9$), lies at about 1\farcs9 from X\#337, thus within the adopted cross-correlation radius. In this paper, we have assumed \cpd7737 to be the only physical counterpart to X\#337.
\\
\item[-]{\bf \cpd7746} was reported as an SB1 system by \citet{GM01}. Located in the outer part of the FOV, it falls on a gap of the \pn\ instrument. The \mos\ count rates are too low to investigate the intra-pointing variability. However, no sign of flaring activities is seen. The spectrum is very poor and the star was not considered further in the present analysis.
\\
\item[-]{\bf \cpd7755} (var.) is an apparently single B1V star. The X-ray flux seems in a higher state during the last three observations than at the beginning of the campaign. An inspection of the images does not reveal any contaminating source. The maximum emission is seen during Obs.~5 and is about a factor 3 to 4 higher than during the first three pointings. We caution however that the count rates are low (between 1 and $4\times10^{-4}$~\cnts\ in the two \mos\ instruments).
\\

\begin{figure}
\centering
\includegraphics[width=\columnwidth]{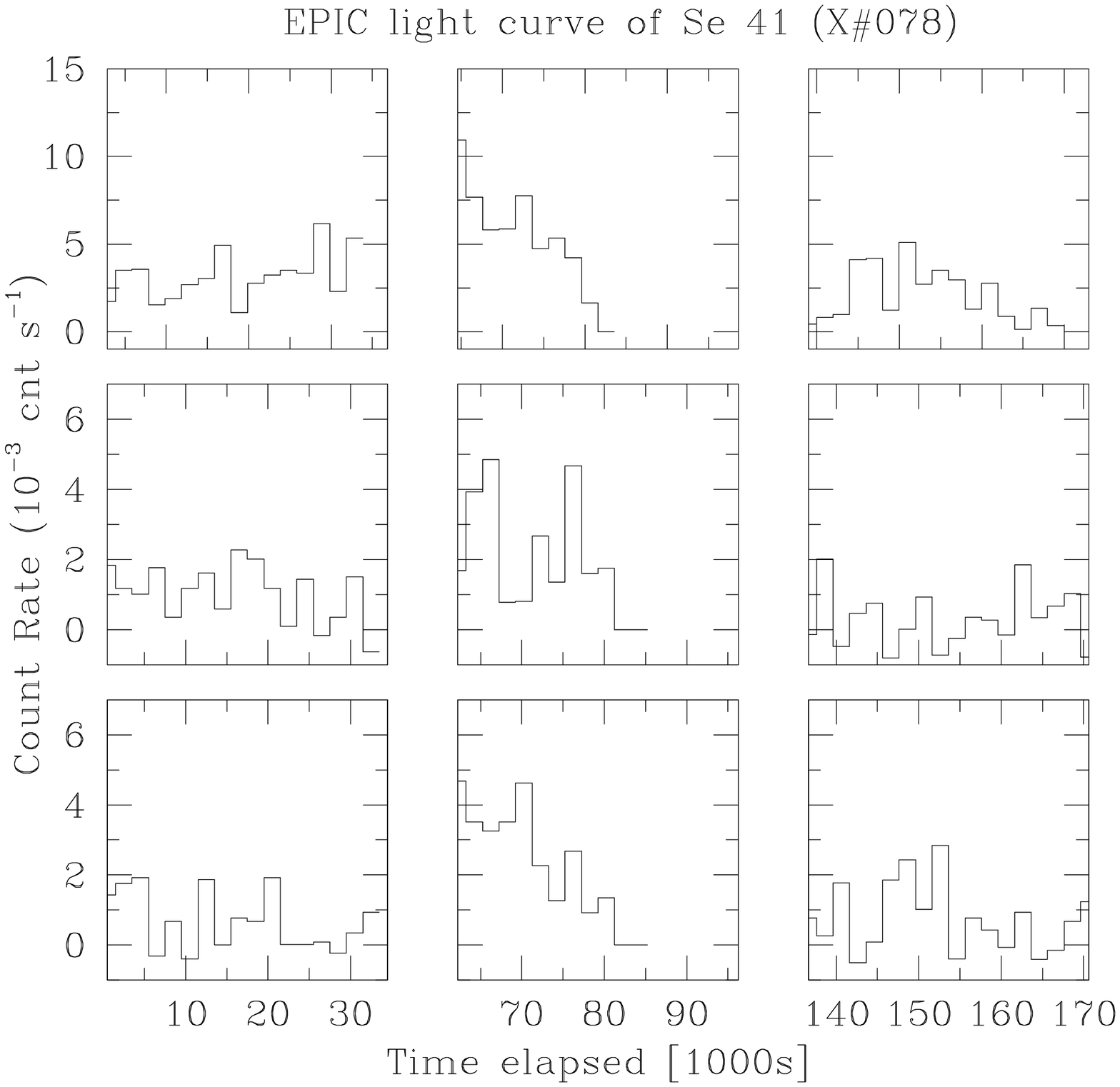}
\caption{{\bf Se~41/SBL~149} \epicpn\ (top panels), \mos2 (middle panels) and \mos1 (lower panels) light curves (in the range 0.5--10.0~keV) during \xmm\ Obs.~1 to 3. The time is given in ks since the beginning of Obs.~1. The bin size is 1~ks. The effective duration of Obs.~2 has been reduced because of a high background event. }
\label{fig: se41}
\end{figure}

\item[-]{\bf \cpd7760} is a visual binary with separation of about 2\arcsec. The A and B components have a probable spectral type B0.5III and B1V respectively. Though the fainter component (\cpd7760B) lies slightly closer to the X-ray source location, the brighter component (\cpd7760A) remains within the adopted 2\farcs5 cross-correlation radius. In Table~\ref{tab: Xob} and subsequent figures, the adopted $L_\mathrm{bol}$ corresponds to the joint contribution of the two stars. Lying in the outer region of the field, outside the \mos\ detection area, this source is only seen in the \pn\ instrument.  Its X-ray emission level is close to our detection limit, but seems to remain constant throughout the six pointings. The obtained spectrum is of poor quality (Fig.~5) and was not considered in the derivation of the \lxlbol\ relations.
\\

\begin{figure}
\centering
\includegraphics[width=\columnwidth]{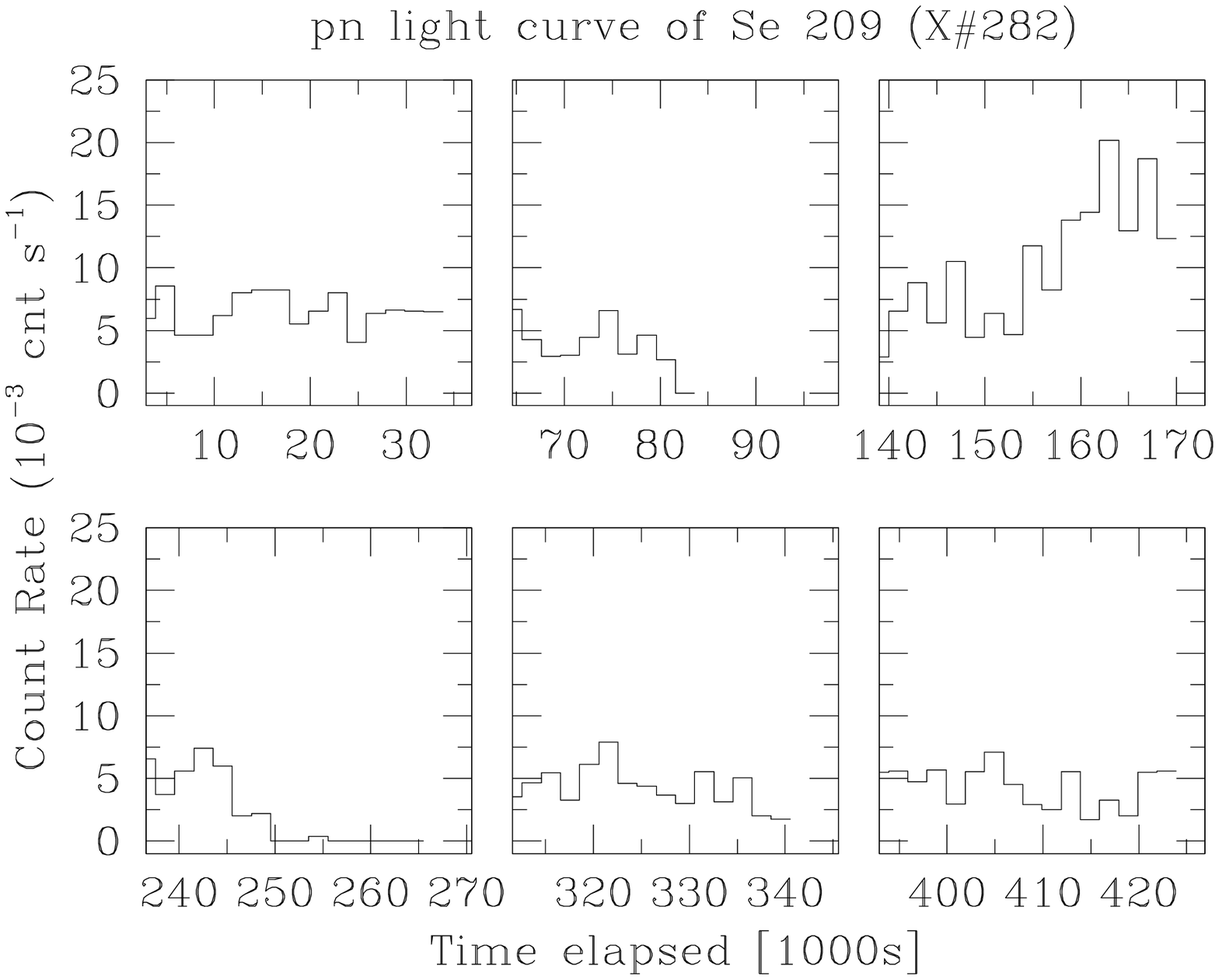}
\caption{{\bf Se~209~/~SBL~394} \epicpn\  light curve (in the range 0.5--10.0~keV) during \xmm\ Obs.~1 to 6. The time is given in ks since the beginning of Obs.~1. The bin size is 2~ks. The effective durations of Obs.~2 and 4 have been reduced because of a high background event.}
\label{fig: se209}
\end{figure}
\item[-]{\bf Se~41~/~SBL~149}  (var.) is a B4V star according to \sbd. With an average \pn\ (vignetting and exposure-time corrected) count rate of about $5\times10^{-3}$~\cnts, it is a relatively bright X-ray source. Its count rate is higher  by a factor 2 to 3 during Obs.~2, whereas it is almost constant during the 5 other exposures. This increase is also seen in the three energy bands. Inspection of the images indicates no sign of a contamination by a nearby flaring source within the extraction region. We therefore consider the flare to be associated with Se~41. We built light curves with a temporal resolution of 1, 2 and 5~ks for all three instruments. They consistently indicate that the emission is decreasing from the beginning of Obs.~2 and reaches a similar  emission level as during the other pointings at the end of Obs.~2 (Fig.~\ref{fig: se41}). 
\\

\item[-]{\bf Se~209~/~SBL~394} (var.) has been proposed either as a binary candidate \citep{Rab96} or as a slowly pulsating B star \citep{ASK01}. It presents an increase of its emission during Obs.~3 (Fig.~\ref{fig: se209}), which yields an average count rate higher by about a factor 2.5 compared to other pointings. 
\\
\item[-]{\bf Se~259~/~SBL~317} is suspected to display radial velocity variations \citep{Rab96} possibly related to an SB2 nature \citep{phd}. This source seems to present a constant, though low, count rate throughout the campaign.
\\
\item[-]{\bf Se~265~/~SBL~200} (var.) is a late B-type main-sequence star (B8.5V according to \sbd) and was quoted as a cluster non-member  by \citet{BVF99}. Its average count rate during Obs.~4 is about a factor two larger than during the other pointings while, actually, its emission level decreases throughout Obs.~4. However, the count rates are quite low ($4\times10^{-3}$~\cnts\ in the \pn\ and $10^{-3}$~\cnts\ in the \mos, on average during the low emission state) and caution should be the rule.
\\

\end{enumerate}


\section{Discussion}\label{sect: discuss}


\subsection{The O-type X-ray emitters}

In Sect.~4, we derived various \lxlbol\  relations corresponding to the O-type stars of our sample. The preferred relation is the scaling law $\log \left(L_\mathrm{X}/L_\mathrm{bol}\right)=-6.912\pm0.153$. Previous works however proposed power law \lxlbol\ relations and we can use the relations quoted in Table \ref{tab: plo} for comparison. The latter ones are indeed formally equivalent to the scaling relations derived in Eqs.~\ref{eq: Oall} to \ref{eq: Om}. 
Working in the 0.2-4.0~keV band with \einst\ data, \citet{SVH90} proposed 
$$\log L_\mathrm{X} = 1.08(+0.06/-0.22) \log L_\mathrm{bol} - 9.38(+2.32/-0.83).$$
 In the 0.1-2.0~keV band, \citet{BSD97} obtained 
$$\log L_\mathrm{X} = 1.13(\pm0.10) \log L_\mathrm{bol} - 11.89 (\pm0.38),$$
 while in the 0.3-12.0~keV band, \citet{ACMM03} reported 
$$\log L_\mathrm{X} = 1.07(\pm0.04) \log L_\mathrm{bol} -  6.2  (\pm0.1)$$
 (though see Sect.~\ref{ssect: carina}).
 Our adopted  slope for a power law \lxlbol\ relation in the \lg\ plane is $0.85\pm0.12$ which is thus significantly lower than the values from \citet{BSD97} and from \citet{ACMM03}. Even our scaling relation, with a slope equal to unity, is still at more than 1-$\sigma$ from the values quoted in the two latter works.
The identification of the strong CWB systems in our sample and their subsequent rejection from the fit can explain part of the observed difference. Indeed including the CWB systems in the fit yields a larger  value for the slope (see Table \ref{tab: plo}).
\\

The direct comparison of the present results with previous works should however be considered with caution because of the different energy bands considered and, in particular, of the lower energy cut-off adopted for the analysis. The influence of the ISM-absorption correction  dramatically increases towards lower energy. For example, assuming an ISM absorption column of $0.26\times10^{22}$~cm$^{-2}$, which is typical for \ngc, the correction factor is  39.3, 2.68, 1.43 and 1.04 respectively in the 0.2-0.5, 0.5-1.0, 1.0-2.5 and 2.5-10.0~keV bands\footnote{These values were computed using the HEASARC's W3PIMMS tool available at {\tt http://heasarc.gsfc.nasa.gov/ Tools/w3pimms.html}. We used a Raymond-Smith model with a temperature of 1.0~keV, but the latter parameter has only a limited influence on the quoted results.}. Hence, a slight uncertainty on the measured fluxes at low energy (or on the estimated ISM absorption) will have a large influence on the intrinsic fluxes computed. The very low energy edge (0.1~keV) adopted by \citet{BSD97} could be one of the reasons for the large dispersion observed in their sample. Another reason is probably the fact that, as explained in \citet{BSC96}, these authors computed the X-ray fluxes by using the count rates together with an energy conversion factor adopted on the basis of the hardness ratio. Though this method is certainly the best that can be done when the number of counts is low, it is inevitably less accurate than the present approach. From Fig.~\ref{fig: lxlb}, we however note that the \citet{BSD97} relation obtained in the 0.1-2.0~keV band for O-type stars remains qualitatively valid in the 0.5-10.0~keV band. Given the previous considerations, this agreement is probably a stroke of good fortune.\\

\subsubsection{The natural dispersion around the canonical \lxlbol\ relation}

From the \rosat\ All-Sky survey data \citep{BSC96}, \citet{BSD97} derived a natural dispersion around the canonical relation of about 0.40 in the \lg\ plane (thus of about a factor 2.5 on the X-ray luminosity), a dispersion that was already much reduced compared to \citet{SVH90} work.  While part of this dispersion is probably due to the non-discrimination of the peculiar objects, it could also reflect the instrumental limitations of the time.  Beyond the intrinsic quality of the data, both the homogeneity of the present sample and the larger value adopted for the low energy cut-off have probably played an important role in preserving a low level of dispersion and, indeed, the dispersion observed around the canonical relation in \ngc\ is definitely very limited.

The current sample of O stars contains both single stars and binaries, and includes objects from different luminosity classes.
However, it  does not extend towards spectral types earlier than O6, nor to bolometric luminosities larger than $\log L_\mathrm{bol}=39.5$. Because of the limited range in $ L_\mathrm{bol}$ (less than 1.5 order of magnitude), extrapolation towards earlier types should be considered with caution. Within these limitations and rejecting the peculiar sources (the two CWBs in the current sample), we note that no systematic discrepancy is observed for stars belonging to different luminosity classes. All the objects (except perhaps the late O-type dwarfs) seem to tightly follow the same canonical relation. This definitely suggests the existence of a common physical mechanism to generate the  X-ray emission for `normal' (i.e. non peculiar) O-type stars.\\

Except for some possible observational artifacts (e.g.\ contamination by neighbouring sources, see Sect.~\ref{sect: indiv}), the sole cause for the observed deviation from the canonical relation in the present sample is X-ray emission from a wind interaction zone. This extra-emission however does not significantly affect all the O-type binaries. It is also the sole identified mechanism that, at our detection limit, yields a modulation of the X-ray flux in our star sample. In consequence, the present analysis suggests that  the intrinsic X-ray emission from non-peculiar O-type stars does not show any significant variability and can be considered as constant for a given star.

\begin{figure}
\includegraphics[width=\columnwidth]{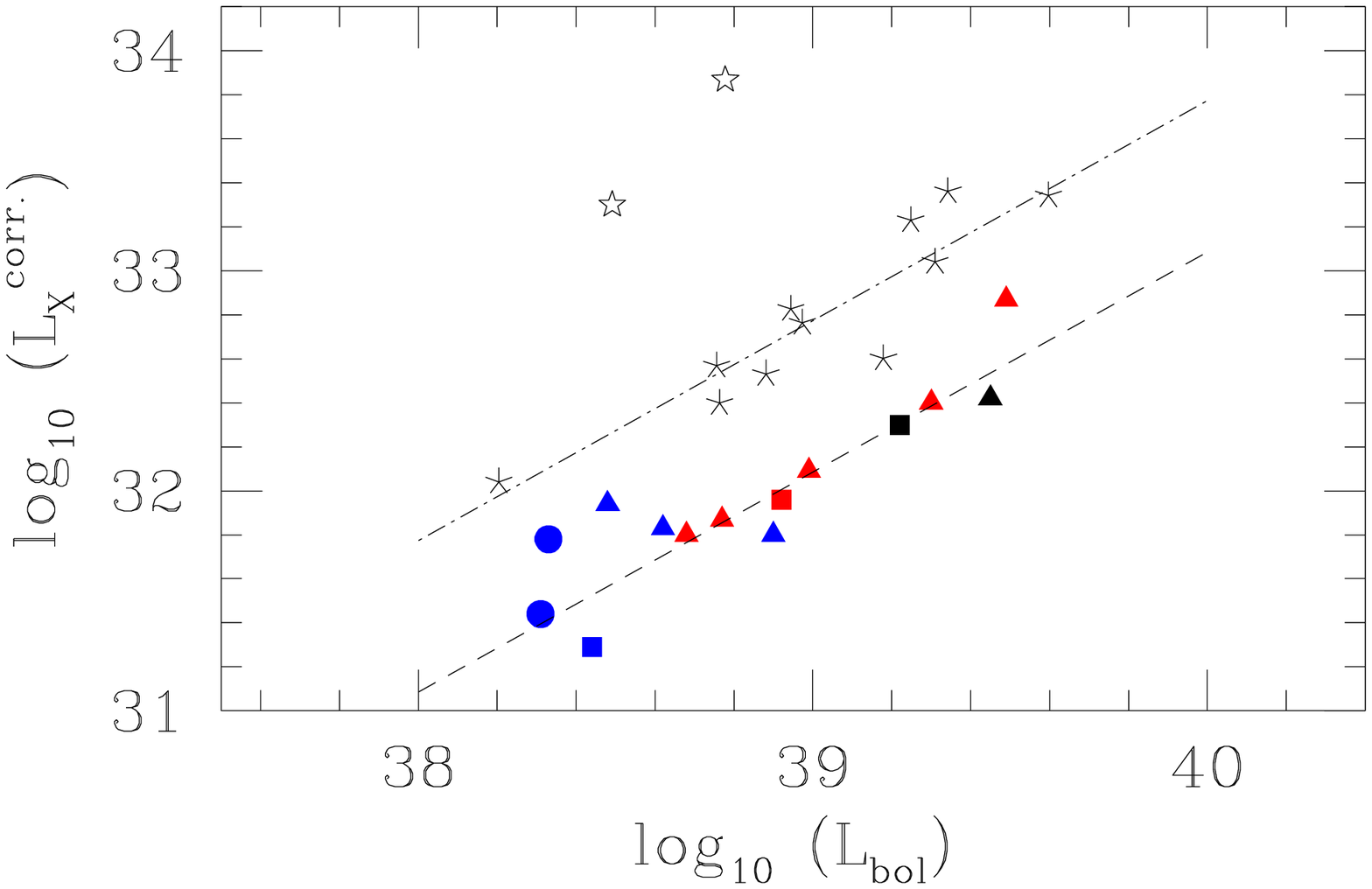}
\caption{Same as Fig.~\ref{fig: clean} for O-type stars. The asterisks give in addition the  \citet{ACMM03} data (obtained in the 0.3-12.0~keV band) and the dashed-dotted line indicates our corresponding best-fit scaling law to these data (Eq.~\ref{eq: slo_acmm}). The open star symbols show  two points that, following  \citeauthor{ACMM03}, were not included in the fit. }
\label{fig: acmm}
\end{figure}

\subsubsection{Constraints from the Carina region} \label{ssect: carina}

\citet{ACMM03} have recently presented a 44~ks \xmm\ observation of the Carina nebula field. As part of their analysis, they reported the canonical relation to be $\log L_\mathrm{X} = 1.07(\pm0.04) \log L_\mathrm{bol} -  6.2  (\pm0.1)$ in the 0.3-12.0~keV domain. Unfortunately, the quoted relation overestimates the X-ray luminosities compared to their actual measurements\footnote{Actually, it is possible that the quoted constant in the power-law is erroneous.}. Using the data from their Table 6 and considering the same objects as they did (11 out of the 13 O stars located in the FOV), we have rederived the best-fit power-law relation. Applying an unweighted linear regression, one obtains :

\begin{figure*}
\includegraphics[width=.33\textwidth]{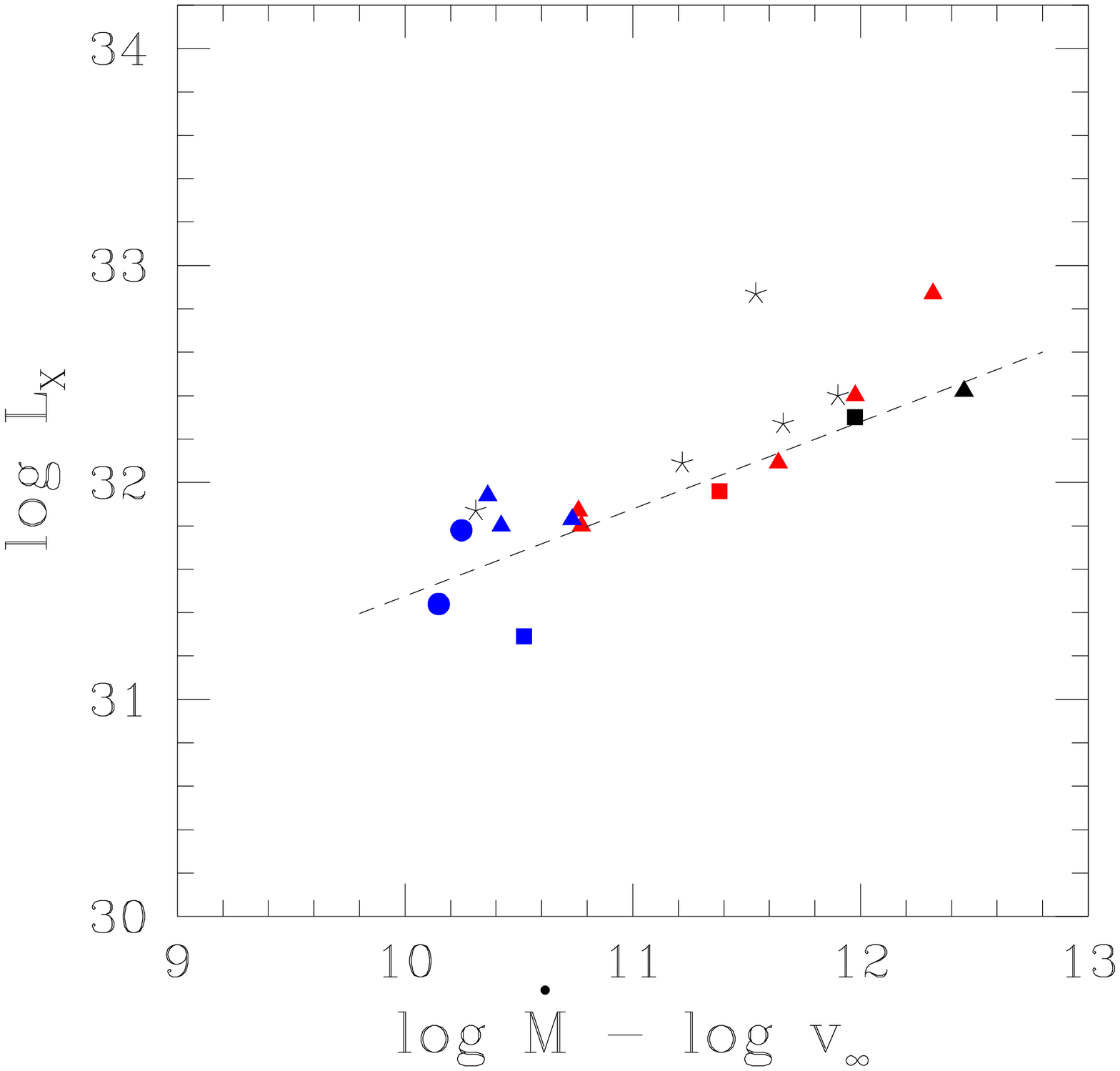}
\includegraphics[width=.33\textwidth]{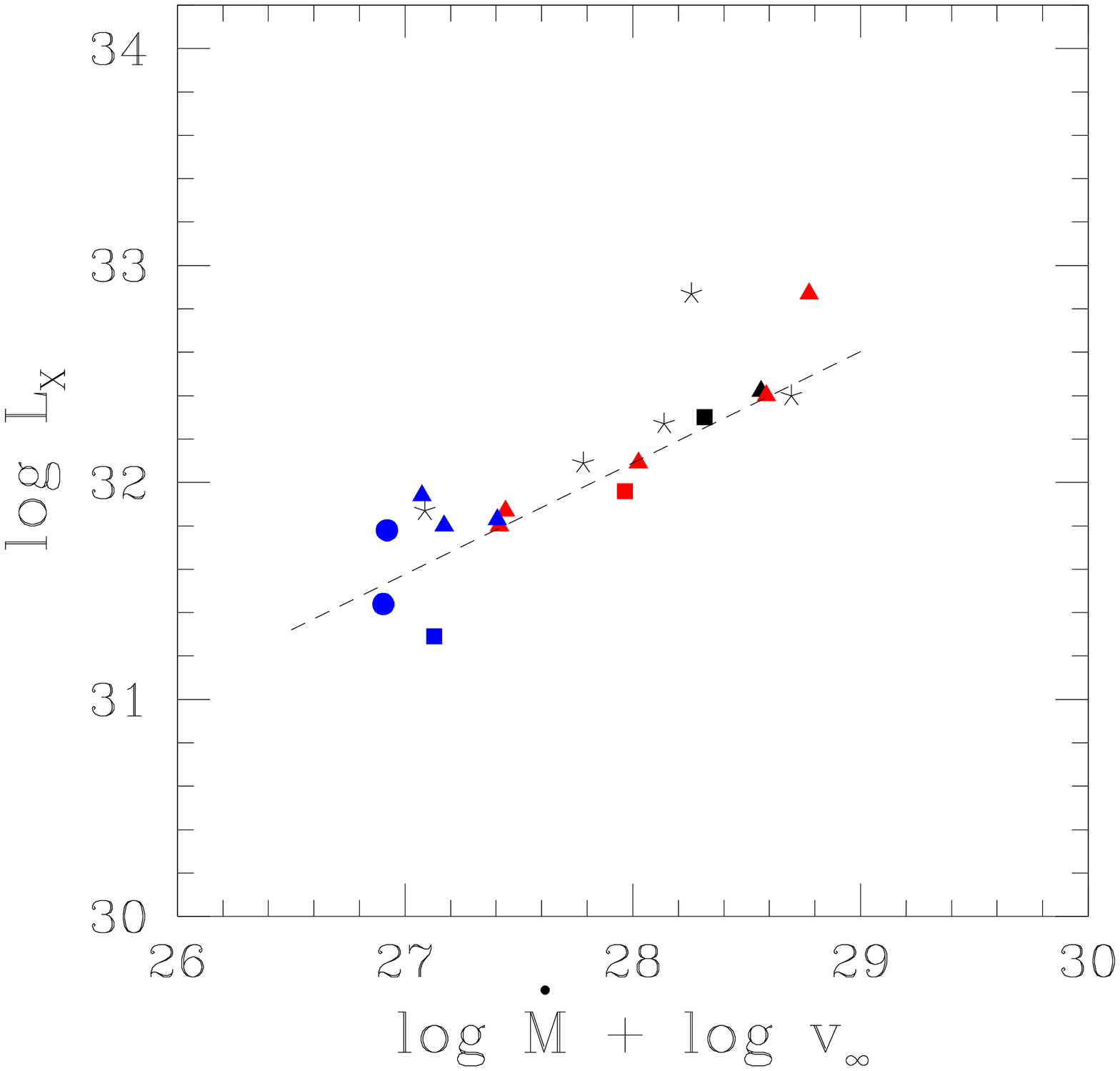}
\includegraphics[width=.33\textwidth]{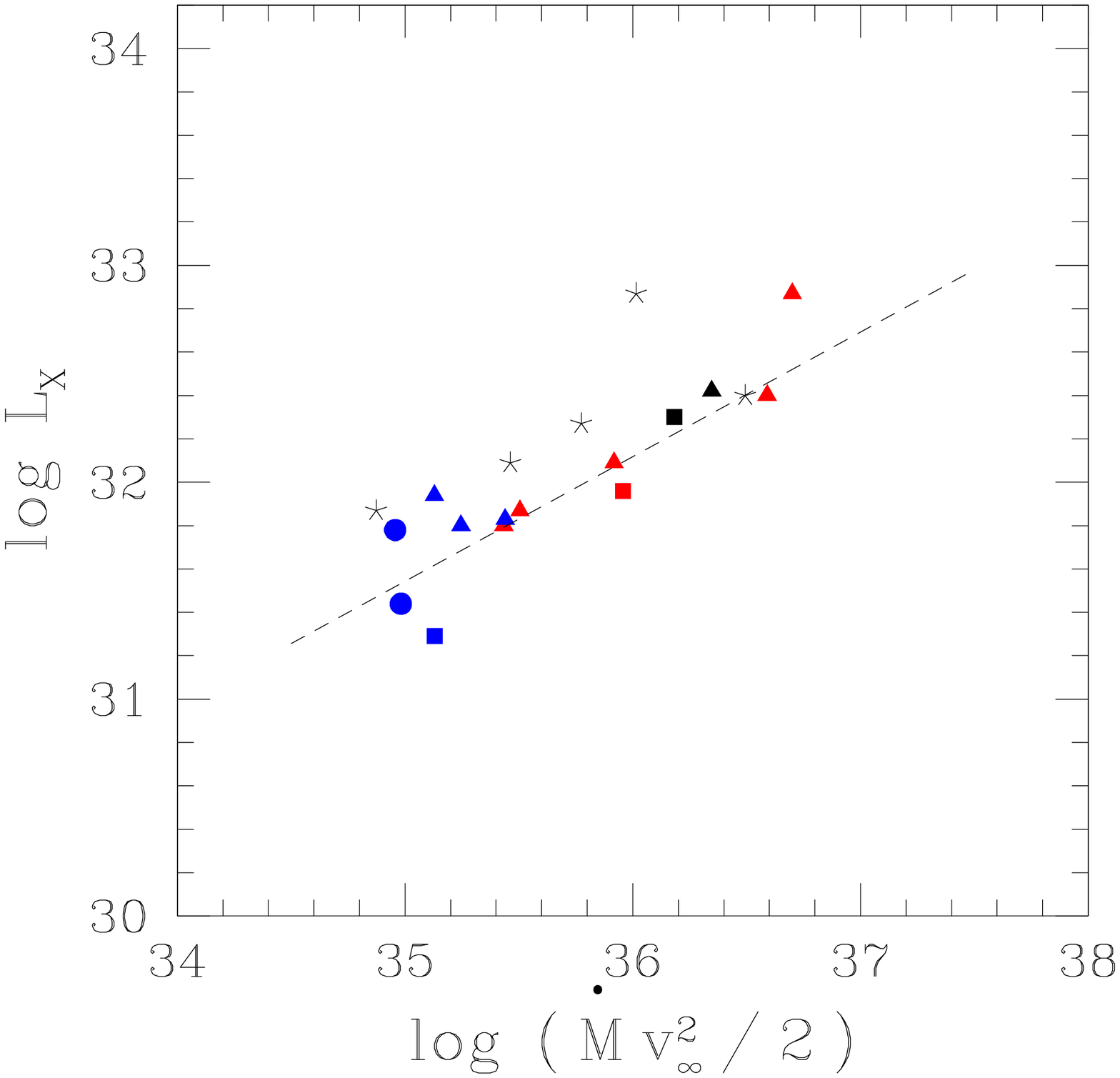}
\caption{ $\log L_\mathrm{X}$ plotted versus different wind parameters. The filled symbols have the same meanings as in Fig.~\ref{fig: lxlb}. Their abscissa were computed using the recipes of \citet{VdKL00, VdKL01}. Dashed lines show the obtained best-fit relations (Eqs.~\ref{eq: OwC_rho} to \ref{eq: OwC_Ek}). The asterisks were computed using the $\dot{M}$ and $v_\infty$ measurements of \citet{HP89} for the available objects. }
\label{fig: OwC}
\end{figure*}

\begin{equation}
\log L_\mathrm{X} = 1.02(\pm0.15) \log L_\mathrm{bol} -  6.9  (\pm5.8) \hspace*{2mm}. \label{eq: plo_acmm}
\end{equation}
The typical dispersion around the relation is about 0.18 in the \lg\ plane, thus very similar to the dispersion obtained from our data set. 
Adopting the  scaling relation model, an unweighted fit  yields 
\begin{equation}
\log \left(L_\mathrm{X}/L_\mathrm{bol} \right) = -6.23\pm0.17  \hspace*{2mm} . \label{eq: slo_acmm}
\end{equation}
 Comparing the latter residuals with those obtained from Eq.~\ref{eq: plo_acmm}, we obtained $F_\chi=0.01$. This clearly indicates that the power law does not provide any significant improvement of the fit compared to the scaling law. The latter relation is plotted in  Fig.~\ref{fig: acmm} together with the data of \citet{ACMM03}.\\

This new analysis of the \citet{ACMM03} data puts into light two important points. First, as in our case, the preferred $\log L_\mathrm{X} - \log L_\mathrm{bol}$ relation takes the form of a scaling law. This clearly contrasts with the previous studies \citep[e.g.][]{ChG91, BSD97} which probably adopted the power law to provide an extra degree of freedom to the fit. The second point relies on the different nature of the considered samples. Our sample contains O6--O9 stars belonging to different luminosity classes whereas \citet{ACMM03} data set is almost uniformly formed by main-sequence stars ranging from O3 to O8.5. However, in both cases, the scaling law is preferred and both samples displayed a limited dispersion around the best-fit relation. We suggest that the slightly larger dispersion observed in \citet{ACMM03} data could come from a less uniform analysis and, in particular, from the handling of the correction for the ISM absorption.  Accounting for the difference in the analysis (different spectral models used) and in the considered energy ranges, the two relations can be regarded as being in mutual agreement if about one fifth of the intrinsic X-ray flux is actually emitted in the 0.3-0.5~keV range. 

\subsubsection{On the origin of the \lxlbol\ scaling}

As pointed out by different authors already, the scaling between \lx\ and \lbol\ could be an indirect effect. Indeed, in the framework of the embedded wind shock model, it is expected that the X-ray luminosity rather scales with the wind parameters or with some physically meaningful combination of them. \citet{SVH90} observed a correlation with the wind momentum ($\dot{M}v_\infty$) and the wind kinetic luminosity ($0.5\dot{M}v^2_\infty$) but did not find a direct correlation with the sole mass-loss rate ($\dot{M}$) nor with the terminal velocity  ($v_\infty$). It is only through the scaling of the properties of the radiatively driven winds with the stellar luminosity that the dependence between \lx\ and \lbol\ is supposed to occur. \citet{KPF96} and \citet{OwC99} later suggested that an important parameter could be the mean wind density, which is related to $\dot{M}/v_\infty$. 
From the exospheric approximation, \citet{OwC99} established that the X-ray luminosity is expected to scale with the wind density parameter $\dot{M}/v_\infty$ in the form  $L_\mathrm{X} \sim (\dot{M}/v_\infty)^2$ for optically thin winds and $L_\mathrm{X} \sim (\dot{M}/v_\infty)^{(1+s)}$ for optically thick winds, with $s$ being the index of the radial power law dependence of the X-ray filling factor $f\sim r^s$. 

 Using the \citet{VdKL00, VdKL01} recipes, we estimated the density parameter ($\dot{M}/v_\infty$), the wind momentum ($\dot{M}v_\infty$) and the wind kinetic luminosity ($0.5\dot{M}v^2_\infty$) for the different O-type objects in our sample. We assumed $v_\infty=2.6\ v_\mathrm{esc}$. Masses and temperatures were taken from the new calibration for O star parameters of \citet{MSH05} while the $L_\mathrm{bol}$ were taken from Sect.~\ref{sect: ET}. Fig.~\ref{fig: OwC} shows the resulting diagrams. Using the same sample of stars as for Eq.~\ref{eq: OmW}, we obtained:
\begin{equation}
 \log L_\mathrm{X}=(0.402\pm0.064) \log \left( \dot{M}/v_\infty \right) + (27.46\pm0.71) \label{eq: OwC_rho}
\end{equation}
\begin{equation}
\log L_\mathrm{X}=(0.513\pm0.075) \log\left(  \dot{M} v_\infty \right) + (17.74\pm2.07) \label{eq: OwC_Pk}
\end{equation}
\begin{equation}
\log L_\mathrm{X}=(0.574\pm0.084) \log \left( 0.5\dot{M} v^2_\infty \right) + (11.45\pm3.01) \label{eq: OwC_Ek}
\end{equation}
with a typical residual dispersion of 0.16. Eq.~\ref{eq: OwC_rho} indicates a filling factor index $s \sim -0.6$.
 
 Wind terminal velocities and star mass-loss rates were measured by \citet{HP89} for five of the brightest stars of our sample. The corresponding observational wind parameters are displayed in Fig.~\ref{fig: OwC}. The agreement with the derived relation is rather acceptable although slight systematic upward shifts are observed. Only \hd248 shows strong deviations from these relations, but the derivations of $\dot{M}$ and $v_\infty$ have probably been biased by the ongoing wind-wind interaction. This clearly supports the idea that the O star X-ray luminosity also scales with various combinations of the wind parameters $\dot{M}$ and $v_\infty$.

However, we emphasize that the canonical relation expressed in Eq.~\ref{eq: OmW} and the Eqs.~\ref{eq: OwC_rho} to \ref{eq: OwC_Ek} are not independent but rather reflect a different point of view on the scaling of the X-ray luminosities with the O-type star properties. Indeed the mass-loss recipes of \citeauthor{VdKL00} analytically expresses $\dot{M}$ in terms of $L_\mathrm{bol}$, $T_\mathrm{eff}$ and the stellar mass $M$. Using  Eq.~\ref{eq: OmW} to express $L_\mathrm{bol}$ in terms of $L_\mathrm{X}$ in the mass loss recipes, we re-derived analytically very similar slopes as those quoted in Eqs.~\ref{eq: OwC_rho} to \ref{eq: OwC_Ek}. For example, subtracting $\log v_\infty$ from $ \log \dot{M}$ almost removes the first order dependency in $T_\mathrm{eff}$.  The remaining relation is dominated by the $L_\mathrm{bol}$ term. From these considerations,  it appears that one can naturally convert the canonical relation between the X-ray and bolometric luminosities into a relation linking the X-ray luminosity  to the wind parameters. For example, $L_\mathrm{X}$ may scales with $\dot{M}/v_\infty$ as proposed by \citeauthor{OwC99} on the basis of more physical considerations.

It is beyond the scope of this paper to pursue the quest for the physical origin of the scaling between the X-ray and the bolometric luminosities. Though the physical processes responsible of the X-ray emission in hot stars are not directly related to $L_\mathrm{bol}$, the scaling between these two parameters is an observational fact. Compared to parameters such as $\dot{M}$ and $v_\infty$, or any combination of these,  the bolometric luminosity is probably much easier to measure and  can be determined with a better accuracy.
We note that this relation is now tightly constrained  and still awaits  to be explained on firm theoretical grounds. 

\subsection{The B-type X-ray emitters}

Among the $\sim$90 B-type stars in the \xmm\ FOV of the present campaign, less than 20\% could be associated with an X-ray counterpart. 
However, except for the earliest sub-types, the B stars are not supposed to emit in the X-ray domain. Their stellar winds are much weaker than those of O-type stars  and should thus not contribute to the X-ray emission. In addition, no coronal emission is expected as these stars are not supposed to have the required convective zones. From our data, either the B-type stars are intrinsic X-ray emitters or the detected X-ray emission is actually associated with a later-type or PMS object along the same line of sight or physically linked to the B star.
The detected dependence between the B star bolometric luminosities and the associated X-ray luminosities, with a linear-correlation coefficient of $\sim$0.75 (thus significant at the 0.01 level) is in favour of the first hypothesis. On the other hand, the fact that we only detect a fraction of the B stars at a given spectral type is clearly against. Indeed, if the X-ray luminosity was actually an intrinsic property of the star, it should be expected that similar stars would  display similar $L_\mathrm{X}$. 

In regard of these two hypotheses, the study of the variability of the individual objects presented in Sect.~\ref{sect: indiv} is particularly relevant. It indicates that, among the 11 B-type stars used to derived the \lxlbol\ relations, four are displaying a flaring activity while two others present long term variations. While flares have been recently observed on a magnetic B-type star \citep[e.g.][]{GrS04}, this remains rather exceptional and such a behaviour is more likely related to the presence of a PMS companion, either in a binary system or located along the line of sight. Indeed,  the X-ray spectra of the ``B-type'' sources in our sample are strikingly similar to those of the numerous PMS stars in the FOV (see our forthcoming analysis for more details about the PMS sources). 

As shown in \citetalias{SGR06}, only about 40 X-ray emitters over the 610 sources are expected to be foreground objects. The possibility for such an object to be located by chance on the same line of sight of a B-type star is thus very limited and can not explain the number of associations between an X-ray source and a B-type star. The density of PMS stars within the \sco/\ngc\ complex is much larger. However,  the limited cross-correlation radius adopted ensures a very low probability ($<0.03$) that such an X-ray emitting PMS star be located by chance on the same line of sight. Therefore, the hidden binary companion becomes a privileged explanation.

While about 50\% of the B-type stars are expected to actually be in binary or multiple systems, typical X-ray emission from a late-type companion is probably too faint to be detected at the distance of \ngc. Hence, only the systems hosting a PMS companion are expected to be seen in the X-ray domain. This could explain the difference between the expected fraction of binaries and the observed detection rate among B stars. In this regard, our conclusions somewhat rejoin those of \citet{SFM05}, who stated that the X-ray emission of weak wind stars is mostly consistent with magnetic activity from known or unseen low-mass companions.

Though the above discussion is clearly in favour of the PMS companion scenario, the observed correlation, at a significance level of 0.01, between  the bolometric luminosity of the B star and the associated X-ray luminosity is intriguing. A possible scenario could imply some connection  between the B-type star and its PMS companion. Though highly putative, such a mechanism could justify the observed link between the B-type properties (\lbol) and the observed X-ray emission from the PMS companion. However, the observed relation might also result from a border effect. Indeed, if the real distribution of the emission `associated' with B stars has about the form of a cluster of points and if one only observes the upper part of the distribution, one might potentially detect an apparent relation between the two considered variables although, actually, it results from an observational cut-off.

\subsection{The demarcation line}

In this paper, we have voluntarily restrained our analysis to objects whose spectra were of sufficient quality to be fitted with 2-T models. This actually led to the rejection of the early B-type stars that were only marginally detected. It is not impossible that some of these actually form the low luminosity tail of the \lxlbol\ relationship derived for O-type stars. Another possibility is that, in this tail, the stellar winds are becoming so weak that only weak shocks are produced, yielding thus a possibly softer emission (see e.g.\ the case of \hd235 in Sect.~\ref{sect: indiv}). Our approach was clearly not designed to investigate the location of the demarcation line. Still, the separation between the two different behaviours clearly occurs at $L_\mathrm{bol}\approx10^{38}$~\ergs. Furthermore, this value clearly separates the O and B-type stars of our sample. 


\section{Conclusions  }\label{sect: ccl}

In this second paper of the series, we have pursued the analysis of the X-ray data concerning the young open cluster \ngc. While \citetalias{SGR06} focused mainly on the  detection and identification of the numerous X-ray sources in the \xmm\ FOV, this paper was devoted to the properties of the rich early-type star population. A detailed census of the OB-type stars within the FOV resulted in more than one hundred objects identified.  Using a limited cross-correlation radius of 2\farcs5, about one third of them could be associated with an X-ray counterpart. Among these, the 15 O-type stars/binaries are all detected in the X-rays as soft and usually bright sources characterized by \mek\ temperatures of k$T=0.3$ and 0.7~keV. On the other hand, the B-type star detection rate only amounts to about 20\%. Compared to the O-type stars, the B-type stars have a similar low energy component  but their second temperature is well above 1~keV. The B-type X-ray emitters are thus significantly harder than the O-type sources. 

O- and B-type emitters clearly present different behaviours in the $\log L_\mathrm{X} - \log L_\mathrm{bol}$ diagram, though both types draw up a linear relation in the log--log plane. The separation between the two sub-sets is located at about $\log L_\mathrm{bol}=38$ (\ergs), as previously suggested by \citet{BSD97}. The dispersion around the expected linear relation is apparently quite small. In the O-type star sample, the two objects that show the largest deviations are known to display an extra-emission component due to a wind interaction (\hda, \citealt{SSG04}; \cpd7742, \citealt{SAR05}). These were thus excluded from the subsequent discussion. We showed that, for the O-type stars, the X-ray luminosities are scaling with the bolometric luminosities. In the 0.5-10.0~keV energy range, we obtained:  
$$\log L_\mathrm{X}-\log L_\mathrm{bol}= -6.912\pm0.153 .$$ 
We also found that a power law relation did not provide any significant improvement to the quality of the fit. The obtained dispersion around this new canonical relation is very limited. It becomes even smaller when excluding the `cooler' (i.e. with a spectral type later than O9) O dwarfs from the fit. In this case, the typical dispersion drops to about 0.087 in the \lg\ plane, thus corresponding to only 20\% on the X-ray luminosities. Within our sample, the only identified mechanism that provides a significant deviation from this relation is extra-emission produced in a wind interaction region. It is also the sole mechanism that, at our detection threshold, produces a significant variability of the observed fluxes in our O-type star sample (see Sect.~\ref{sect: indiv}). Though relatively limited, the present sample suggests thus that the intrinsic X-ray emission from O-type stars is very tightly correlated with their bolometric luminosity. Beyond the two strong CWB systems, our sample is formed by single stars and binaries belonging to different luminosity classes. Though not extending towards sub-types earlier than O6, they all seem to follow the canonical relation, suggesting thus a common mechanism for X-ray production in these objects.

We also provide a new analysis of recent flux measurements obtained in the Carina region \citep{ACMM03}. This sample is mainly formed by main-sequence stars ranging from O3 to O8.5. We note that this new analysis confirms much of our present conclusions. The best \lxlbol\ relation is indeed in the form of a scaling law rather than a power law. The dispersion around the obtained relation is very limited and the difference with our own relation might possibly be accounted for by the different energy ranges considered.\\

We emphasize that this apparent scaling might indirectly result from a scaling of the X-ray luminosity with the wind properties, themselves scaling with the bolometric luminosities for these stars with radiatively driven winds. Clues for this are provided by the work of \citet{VdKL00, VdKL01} which has allowed us to naturally convert the scaling of the X-ray luminosity with the bolometric luminosity into a scaling with the wind parameters.
Nonetheless the \lxlbol\ relation probably remains the most accurate observational constraint to link the intrinsic X-ray emission of the O-type stars with their fundamental properties.

Turning to B-type stars, the fact that only about one quarter of the stars of a given spectral sub-type are actually associated with an X-ray source argues strongly against X-ray emission being  an intrinsic property of these stars. We however note that we still observed a linear relation that links $\log L_\mathrm{X}$ and $\log L_\mathrm{bol}$: 
$$\log L_\mathrm{X} = (0.22\pm0.06) \log L_\mathrm{bol} + 22.8 (\pm 2.4) $$
The dispersion around this relation is quite limited ($\sim$0.14) and the linear pattern is also seen in the different energy sub-ranges considered. From our analysis, the most probable explanation points towards the X-ray emission originating from a low mass PMS physical companion. The observed relation between \lx\ and \lbol\ remains however a puzzle and could eventually result from an observational effect.  Alternatively it could be linked to some particularities of the B-type stars in \ngc\ or to a putative interaction between the PMS object and its B-type companion, yielding an X-ray emission partly governed by the intrinsic properties of the B-type primary.  \\

Finally, we note that the separation line between the O- and ``B-type'' behaviours (around $L_\mathrm{bol}=10^{38}$~\ergs) is poorly mapped by our present sample. The adopted criterion to preserve the homogeneity of the spectral analysis (requiring at least 2-T models) has led to  the rejection  of the objects in the transition zone. The extent of the canonical relation towards lower luminosities probably deserves a more particular attention. Dedicated observations, combined with the already observed fields, could help to increase the number of objects in this zone. This is probably a necessary condition to answer with more details this still open question.


\section*{Acknowledgments}
The authors are greatly indebted to the `Fonds National de la Recherche Scientifique` (FNRS), Belgium, for multiple supports. This research is supported in part by contract P5/36 ``P\^ole d'Attraction Interuniversitaire'' (Belgian Federal Science Policy Office) and through the PRODEX XMM and INTEGRAL contracts. 
The SIMBAD and WEBDA databases and the Vizier catalogue access tool (CDS, Strasbourg, France) have been consulted for the bibliography and for the purpose of object cross-identification. 


\bibliographystyle{mn2e}
\bibliography{/datas6/XMM_CAT_PAPER/ngc6231_Xcat}

\bsp
\label{lastpage}

\end{document}